\documentclass[10pt,a4paper]{article}
 
\usepackage[latin1]{inputenc}
\usepackage{float}
\usepackage{bm}
\usepackage{bbm}
\usepackage{amsfonts}
\usepackage{amsthm}
\usepackage{amsmath}
\usepackage{mathtools}
\usepackage{mathrsfs}

\renewcommand{\[}{\begin{equation}\begin{aligned}}
\renewcommand{\]}{\end{aligned}\end{equation}}

\usepackage{slashed}

\usepackage{cite}
\usepackage{bbold}
\usepackage{todonotes}
\usepackage{graphics}
\usepackage{graphicx}
\usepackage{tikz}
\usetikzlibrary{patterns}
\usetikzlibrary{calc,decorations.markings,arrows.meta,shapes.misc,decorations.pathmorphing,bending}
\tikzset{
  branch point/.style={cross out,draw=black,fill=none,minimum size=2*(#1-\pgflinewidth),inner sep=0pt,outer sep=0pt}, 
  branch point/.default=5
}
\tikzset{
  branch cut/.style={
    decorate,decoration=snake,
    to path={
      (\tikztostart) -- (\tikztotarget) \tikztonodes
    },
    execute at begin to={{
      \coordinate (A) at ($(\tikztostart)!.8!-10:(\tikztotarget)$);
      \coordinate (B) at ($(\tikztostart)!.8!10:(\tikztotarget)$);
      \coordinate (AB/3) at ($(A)!1/3!(B)$);
      \coordinate (2AB/3) at ($(A)!2/3!(B)$);
      \coordinate (C) at ($(AB/3)!2/(3*sqrt(3))!-90:(B)$);
      \coordinate (D) at ($(2AB/3)!4/(3*sqrt(3))!-90:(B)$);
    }}
  }
}

\newcommand{\beq}{\begin{equation}}
\newcommand{\eeq}{\end{equation}}
\newcommand{\bea}{\begin{eqnarray}}
\newcommand{\eea}{\end{eqnarray}}
\newcommand{\nn}{\nonumber}

\def\lsi{\raise0.3ex\hbox{$<$\kern-0.75em\raise-1.1ex\hbox{$\sim$}}}
\def\gsi{\raise0.3ex\hbox{$>$\kern-0.75em\raise-1.1ex\hbox{$\sim$}}}

\newcommand\brwrap[3]{%
  \setbox0=\hbox{$#2$}
  \left#1\vbox to \the\ht0{\hbox to 0pt{}}\right.\kern-.2em
  \begingroup #2\endgroup\kern-.15em
  \left.\vbox to \the\ht0{\hbox to 0pt{}}\right#3
}

\def\bra#1{\langle #1|}
\def\ket#1{|#1 \rangle}

\usepackage{setspace}
\usepackage{fancyhdr}
\usepackage{listings}
\usepackage{changepage}
\usepackage[colorlinks=true, urlcolor=black, plainpages=false, pdfpagelabels]{hyperref}
\usepackage{etoolbox}
\usepackage{amssymb}
\usepackage[parfill]{parskip}
\title{{\bf Compton scattering from superstrings}}

\author{Thales Azevedo$^{a}$\footnote{\href{mailto:: thales@if.ufrj.br}{thales@if.ufrj.br}}~,
Daniel E. A. Matamoros$^{b}$\footnote{\href{mailto:: dea23@fsu.edu }{dea23@fsu.edu }}~, 
and Gabriel~Menezes$^{c}$\footnote{\href{mailto:: gabriel.menezes10@unesp.br}{gabriel.menezes10@unesp.br} (On leave of absence from Departamento de F\'{i}sica, Universidade Federal Rural do Rio de Janeiro.)} \vspace{14pt} 
\\
\it $^{a}$Instituto de F\'isica, Universidade Federal do Rio de Janeiro, 
\\\vspace{8pt} 
Av. Athos da Silveira Ramos 149, Rio de Janeiro, RJ 21941-972, Brazil\\
\it $^{b}$Physics Department, Florida State University,
\\\vspace{8pt} 
\it  Tallahassee, Florida 32306-4350, USA
\\
\it $^{c}$Instituto de F\'isica Te\'orica, Universidade Estadual Paulista,
\\ \it Rua Dr.~Bento Teobaldo Ferraz, 271 - Bloco II,  S\~ao Paulo, SP 01140-070, Brazil}

\begin{document}
\maketitle

\begin{abstract}

We propose a candidate Compton amplitude which is valid for any (integer) quantum spin and free from any spurious poles. We consider the cases of electromagnetism and gravity. We obtain such amplitudes by calculating the corresponding ones from superstring theory involving states on the leading Regge trajectory. To extract the associated field-theory amplitudes a few considerations in the form of simple physical constraints are required, such as: Soft momentum transfer, compactification of polarizations and consistent factorization in the physical channels. We believe the present exploration will be significantly relevant for the physics of compact binary systems with spin.

\end{abstract}


\section{Introduction}

An important insight arising from contemporary studies of quantum scattering amplitudes is that gravitational amplitudes are simpler than one would expect from a brute-force Feynman diagram calculation~\cite{Bern:2022wqg}. The double copy prescription determines that the gravitational amplitudes are obtained as a product of two Yang-Mills amplitudes. The original formulation by Kawai, Lewellen and Tye states that a closed string amplitude at the tree level is given by a sum over the products of two open string amplitudes at tree level~\cite{KLT}. More recently, Bern, Carrasco and Johansson demonstrated that gravitational numerators are, in a sense, the square of the kinematic numerators of the Yang-Mills theory~\cite{BCJ1,BCJ2,Bern:2019prr,Brandhuber:2021kpo}, which has revealed another aspect of the double copy. This approach also connects classical solutions of Yang-Mills theory and gravity. In particular, point charges in Yang-Mills theory map to point sources in gravity~\cite{89,90,91,92}. There are also indications that the double copy may encompass bound states and particles with spin~\cite{96,97}, and should also be relevant for higher-derivative gravity theories~\cite{Johansson:2017srf,Johansson:2018ues,Menezes:2021dyp,Menezes:2022jow,Lescano:2023pai,Azevedo:2018dgo,Azevedo:2019zbn}.

This current trend can be expected to be of extreme importance for the physics of gravitational waves. Indeed, the broader interest in this program is to scrutinize the application of modern scattering amplitude techniques to address the two-body gravity problem. This question arises naturally due to the birth of gravitational wave astronomy, driven by landmark observations made by the LIGO and Virgo collaborations~\cite{LIGO1,LIGO2,LIGO3,LIGO4,LIGO5}. Such techniques complement several methods presented in the literature and have been of utmost significance in fostering interdisciplinary collaborations, including the exploration of new research directions.  Amplitudes have already been successfully applied to understand aspects of the general relativistic two-body problem~\cite{Bjerrum-Bohr:2022blt,Kosower:2022yvp,Kosower:19,Cachazo:2017jef,Guevara:2017csg,Sturani:2021ucg,Neill:2013wsa,Bjerrum-Bohr:14,Bjerrum-Bohr:2014lea,Bjerrum-Bohr:2014zsa,Bjerrum-Bohr:16,76,Bjerrum-Bohr:2021vuf,Bjerrum-Bohr:2021din, Herrmann:2021tct,Cristofoli:2021vyo, Bern:2021dqo,Bern:2021yeh,Herrmann:2021lqe, DiVecchia:2021bdo, Bern:2020gjj,Moynihan:2020gxj,Cristofoli:2020uzm,Parra-Martinez:2020dzs,Haddad:2020tvs,AccettulliHuber:2020oou,Moynihan:2020ejh,Manu:2020zxl,Sahoo:2020ryf,delaCruz:2020bbn,Bonocore:2020xuj,Mogull:2020sak,Emond:2020lwi,Cheung:2020gbf,Mougiakakos:2020laz,Carrasco:2020ywq,Kim:2020cvf,Bjerrum-Bohr:2020syg,Gonzo:2020xza,delaCruz:2020cpc,Cristofoli:2021jas,Bautista:2021llr,  Brandhuber:2021eyq,Brandhuber:2021bsf,Brandhuber:2021kpo,Aoude:2021oqj,Cho:2022syn,Bern:2020buy,Bautista:2021wfy,yutinspin,Alessio:2022kwv,Bern:2022kto,FebresCordero:2022jts,Bohnenblust:2023qmy,Maybee:2019jus}. In fact, in recent years there has been a growing interest in amplitude calculations applied to classical gravity. And the benefit is not only on the technical side, but also in obtaining a deeper knowledge that field theory provides for long-standing problems such as
interpretation of infrared divergences \cite{Foffa:2019yfl} and the connection between
effective potential and scattering angle~\cite{Bern:2019nnu,Bini:2019nra,Bini:2020wpo}.

The use of effective field theory methods to classical gravity were
investigated in a pioneering way by \cite{18}, see also~\cite{19,Iwasaki:1971vb,HariDass:1980tq,Damour:1995kt,77,78,Foffa:2019yfl,Kalin:2020mvi,Kalin:2020fhe}. They have been successfully applied to the general problem of two-body relativistic dynamics in the post-Newtonian (PN) approximation up to the fourth perturbative order, with partial results in the fifth perturbative order~\cite{Foffa:2021pkg,Almeida:2021xwn}. The fourth Post-Minkowskian (PM) Order was studied in Ref.~\cite{Dlapa:2024cje}, and, within amplitude-based methods, in Refs.~\cite{Bern:2021dqo,Bern:2021yeh}. In particular, in~\cite{Foffa:2016rgu} the connection between the classical two-body gravitational problem and the calculation of quantum field theory amplitudes was highlighted, showing how the evaluation of the effective potential of $n$-th order in PN approximation can be interpreted in terms of a calculation of quantum self-energy in diagrams of massless particles with multiple self-interactions. Solving the equations iteratively in perturbative orders, as usually done in PN approximation, is reminiscent of the application of perturbative methods in quantum field theory, where scattering amplitude calculations are traditionally organized in Feynman diagrams.

In this context, we have witnessed a particularly intense flurry of activities about the gravitational interaction of spinning particles~\cite{worldsheetpaper,Li:2018qap,Goldberger:2017ogt,Bjerrum-Bohr:2020syg,Bonocore:2020xuj,Bautista:2019evw,Bautista:2021wfy,Bautista:2022wjf,Alessio:2022kwv,Menezes:2022tcs,Aoude:2022trd,Aoude:2021oqj,Haddad:2023ylx,Chiodaroli:2021eug,Chung:19,Cangemi:2023ysz,Cangemi:2023bpe,Bjerrum-Bohr:2023jau,Bjerrum-Bohr:2023iey,yutinspin2,yutinspin,mogullspin2,Guevara:2019fsj,Guevara:19a,Arkani-Hamed:20,Damour:2024mzo,Bern:2023ity,Bern:2020buy,Bern:2022kto,Maybee:2019jus,FebresCordero:2022jts,Bohnenblust:2023qmy,Burger:2019wkq,Levi:2020lfn,Levi:2020uwu,Liu:2021zxr,Cho:2022syn,Faye:2006gx,Blanchet:2006gy,Damour:2007nc,Vines:2017hyw,Bini:2017xzy,Bini:2018ywr,Scheopner:2023rzp,Brandhuber:2023hhl,DeAngelis:2023lvf}~\footnote{For a much larger body of older and current research please see the
references within~\cite{Li:2018qap,Goldberger:2017ogt,Bjerrum-Bohr:2020syg,Bonocore:2020xuj,Bautista:2019evw,Bautista:2021wfy,Bautista:2022wjf,Alessio:2022kwv,Menezes:2022tcs,Aoude:2022trd,Aoude:2021oqj,Haddad:2023ylx,Chiodaroli:2021eug,Chung:19,Cangemi:2023ysz,Cangemi:2023bpe,Bjerrum-Bohr:2023jau,Bjerrum-Bohr:2023iey,yutinspin2,yutinspin,mogullspin2,Guevara:2019fsj,Guevara:19a,Arkani-Hamed:20,Damour:2024mzo,Bern:2023ity,Bern:2020buy,Bern:2022kto,Maybee:2019jus,FebresCordero:2022jts,Bohnenblust:2023qmy,Burger:2019wkq,Levi:2020lfn,Levi:2020uwu,Liu:2021zxr,Cho:2022syn,Faye:2006gx,Blanchet:2006gy,Damour:2007nc,Vines:2017hyw,Bini:2017xzy,Bini:2018ywr,Scheopner:2023rzp,Brandhuber:2023hhl,DeAngelis:2023lvf}.}. Indeed, gravitational dynamics of spinning objects has been a subject of vigorous study for several years~\cite{Mathisson:1937zz,Papapetrou:1951pa,Pirani:1956tn,Tulczyjew:59} in both PN and PM approaches. Since scattering amplitudes are envisaged as useful means to incorporate black-hole dynamics, the expectation is that gravitational dynamics of Kerr black holes can be encoded in the dynamics described by a set of higher-dimension operators with presumably known (or to be inferred) Wilson coefficients. Early work suggested that the amplitudes involving massive low-spin states and gravitons could be used to calculate the lowest orders in multipolar spin expansion from black-hole observables, such as scattering angles and gravitational potentials. In fact, conservative observables for Kerr black holes have been studied in a variety of different frameworks, including higher-spin effective field theories and worldline field theories~\cite{Bern:2020buy,Liu:2021zxr,Jakobsen:2021lvp,Jakobsen:2022fcj,Riva:2022fru}. 

An outstanding problem emerging in this regard is the evaluation of the gravitational Compton scattering amplitude -- the opposite-helicity version is a key ingredient in the evaluation of the conservative potential of compact binary systems with spin~\cite{Bern:2023ity,Bern:2020buy,Bern:2022kto,Aoude:2022thd,Aoude:2023vdk} and also waveform calculations with spin~\cite{DeAngelis:2023lvf,Brandhuber:2023hhl,Bautista:2021inx,Aoude:2023dui}. Previous calculations in the literature have shown how challenging this task can be~\cite{Arkani-Hamed:2017jhn,Johansson:19}~\footnote{Opposite-helicity Compton amplitude built from BCFW recursion relations have unphysical poles because this procedure is no longer predictive in the case of higher spins -- the product of BCFW shifted three-point amplitudes does not approach zero fast enough as the shift parameter goes to infinity~\cite{Aoude:2022trd}. Nevertheless, we remark that if one starts with a local Lagrangian the amplitudes calculated from it using standard Feynman rules are expected to be manifestly local; see Ref.~\cite{Bern:2023ity} for a calculation for the case of massive spin 3 electromagnetic Compton.}. In any case, subsequent research efforts have successfully navigated initial obstructions and by now important ongoing research is being conducted to address this problem~\cite{Chiodaroli:2021eug,Aoude:2022trd,Bautista:2021wfy,Bautista:2022wjf,Haddad:2023ylx,Chung:19,Cangemi:2023ysz,Cangemi:2023bpe,Bjerrum-Bohr:2023jau,Bjerrum-Bohr:2023iey,Scheopner:2023rzp,Skvortsov:2023jbn,Cangemi:2022bew}.

This scenario also opens up a great opportunity to investigate string amplitudes. Indeed, one of the most notable features of string theory is the presence of an infinite tower of massive spinning modes in its spectrum, required for consistency. Moreover, the string theory framework allows one to consider the scattering of states of indefinite spin in a systematic manner~\cite{Schlotterer:2010kk}.
 However, some difficulties have been highlighted in the literature~\cite{Cangemi:2022abk,Giannakis:1998wi,Bern:2020buy}. Nevertheless, as string theory provides a consistent theory of massive interacting higher-spin particles, one can regard it as a natural framework to look into classical objects which involve spin-to-infinity limits.

 Indeed, in this paper we wish to report progress in that direction. First, as a promissing result, we extract from a string-theory based calculation the correct tree-level Compton amplitude for a Kerr black hole for quantum spin two~\footnote{Paolo di Vecchia and Francesco Alessio have also obtained the same result through a slightly different calculation. We thank them for sharing their unpublished notes with us.}. 
Then, having succeeded in recovering this result, we propose a candidate Compton amplitude which is free from any spurious poles and valid to any integer quantum spin.

In summary, our claim is: string-theory amplitudes could potentially contain amplitudes that
describe compact astrophysical bodies, in the form of specific contributions (among many others) in disguise inside the full amplitude. To extract those hidden contributions, all other contributions must be
projected out, and we propose that this can be achieved by imposing the following three constraints, to be explained in detail in this paper:

\begin{itemize}

\item soft momentum transfer;

\item compactification of polarizations, so that the mass level matches the spin of the leading Regge states;

\item consistent factorization in the physical channels for any helicity choice.

\end{itemize}

Perhaps our procedure is best justified a posteriori, given the results we obtain. With the simple rules above we manage to obtain sensible field-theory Compton amplitudes which are physically sound and whose classical limit agrees with that of Ref.~\cite{Cangemi:2023ysz} up to fifth order in the ring radius (in the gravitational case).
In fact, we argue that such a reasoning can also be used to extract the correct 3-point root-Kerr amplitude from the corresponding superstring amplitude~\cite{Schlotterer:2010kk,Cangemi:2022abk}, up to terms of order (mass)${}^{-1}$. Throughout the paper, we employ natural units, with $\hbar = c = 1$.

\section{Compton scattering for open and closed strings}

\subsection{Open strings}

Let us calculate the Compton amplitude involving two photons and two massive spin-$S$ particles. For simplicity in this paper we will only study leading Regge states of the NS sector of the superstring. We consider the Minkowski metric as $\eta_{\mu\nu} = \textrm{diag}(-,+,+,\cdots,+)$ and Mandelstam variables defined as $s_{ij} = - (k_i + k_j)^2$, with $s_{12} = s$, $s_{13} = t$ and $s_{14} = u$. In addition, $k_1^2 = k_3^2 = - m^2$. Here $s+t+u = 2 m^2 = 2(S-1)/\alpha^{\prime}$, and the leading Regge state satisfies $m^2 \alpha^{\prime} = n = S-1$. The vertex operators required here are the following~\cite{Schlotterer:2010kk}
\beq
V_n^{(-1)}(k;z) =  \frac{e}{(2 \alpha^{\prime})^{n/2}}
\epsilon_{\mu_1 \ldots \mu_n \nu}(k) : \left( \prod_{j=1}^{n} i \partial X^{\mu_j} \right)  \psi^{\nu}(z) e^{-\phi(z)} 
e^{i k \cdot X(z)} :
\eeq
and
\beq
V_0^{(0)}(q;z) = e \sqrt{2 \alpha^{\prime}} \epsilon_{\nu}(q) : \left( (q \cdot \psi) \psi^{\nu}(z) 
+ \frac{1}{2 \alpha^{\prime}} i \partial X^{\nu} \right) e^{i q \cdot X(z)} : .
\eeq
See also Ref.~\cite{Sagnotti:2010at} for the vertex operators of the open bosonic string. We employ a given representation of the massive polarization tensors $\epsilon_{(S)}$ (they are symmetric, traceless and transverse in all indices) of spin-S particles in which they are written as symmetric tensor products of massive spin-$1$ polarization vectors $\epsilon_1$, namely 
$\epsilon_{(S)} = \bigotimes_{S} \epsilon_1$. We should also consider $V_0^{(0)}$ -- in order to cancel the background ghost charge on a genus $g$ Riemann surface, a $g$-loop superstring amplitude needs an overall superghost charge of $2g-2$. Finally, we will take $V_0^{(0)}$ to represent the photon operator in the zero picture. Our disk amplitude on the leading Regge trajectory reads 
\beq
A_4({\bf 1}, 2^{h_2}, {\bf 3},4^{h_4}) = e^{-\lambda} \int_{-\infty}^{\infty} dz_3 
\left\langle c(z_1) V^{(-1)}_n(k_1;z_1) \, c(z_2) V_0^{(0)}(q_2;z_2) 
\, V_n^{(-1)}(k_3;z_3) \, c(z_4) V_0^{(0)}(q_4;z_4) \right\rangle 
\eeq
where the factor $e^{-\lambda}$ comes from the Euler-number term in the action. The above bold notation refers to massive spinning particles. For convenience we write the amplitude as
\bea
A_4({\bf 1}, 2^{h_2}, {\bf 3},4^{h_4}) &=& \epsilon_{\alpha}(2) \epsilon_{\beta}(4) \,
\epsilon({\bf 3}) \cdot A^{\alpha\beta}_4({\bf 1}, 2^{h_2}, {\bf 3},4^{h_4})  \cdot \epsilon({\bf 1})
\nn\\
&=& \epsilon_{\alpha}(2) \epsilon_{\beta}(4) 
\left( A^{\alpha\beta}_4({\bf 1}, 2^{h_2}, {\bf 3},4^{h_4}) \right)^{a_1 \ldots a_n a_S}\,_{c_1 \ldots c_n c_S}
\, \,  \epsilon_{a_1 \ldots a_n a_S}({\bf 1}) \epsilon^{c_1 \ldots c_n c_S}({\bf 3})
\eea
where the dot products refer to the contraction of the higher-spin indices for legs 1 and 3. Henceforth we are suppressing the higher-spin indices for clarity. Moreover, we take $z_1 \to \infty$, $z_2=1$ and $z_4 = 0$, which implies that $z_3 \in [0,1]$. 

We proceed to calculate the integrals by employing well known correlators in the literature~\cite{Schlotterer:2010kk,Agia:2023lfl}. There are associated constants of proportionality that are fixed by requiring that the amplitude describes the interaction of photons with massive particles. Hence a straightforward calculation produces the following color-ordered amplitude
\beq
A^{\alpha\beta}_4({\bf 1}, 2^{h_2}, {\bf 3},4^{h_4}) \equiv A^{\alpha\beta}_t({\bf 1}, 2^{h_2}, {\bf 3},4^{h_4}) = 
\frac{{\cal V}_t}{\alpha^{\prime} t \, \prod_{k=1}^{2n-1} (\alpha^{\prime} t - k)}
{\cal A}_0^{\alpha\beta},
\eeq
where ${\cal V}_t \equiv \Gamma (1-s \alpha^{\prime})\Gamma (1-u \alpha^{\prime})/
\Gamma(1+\alpha^{\prime} t - 2n)$. The explicit expression for ${\cal A}_0^{\alpha\beta}$ (which does not contain propagators) is rather involved and we refer the reader to the appendix for details, as well as for some explanation concerning notation. It is easy to see that the full amplitude is invariant under the simultaneous exchange of  $\alpha \leftrightarrow \beta$ and $q_2 \leftrightarrow q_4$.

\subsection{Closed strings}

In order to display the Compton amplitude for the closed string, we will resort to the known KLT relations~\cite{Bjerrum-Bohr:2010pnr} ($N$ is the number of external particles)
\bea
M_N &=& \left( \frac{-i}{4} \right)^{N-3}
\sum_{\sigma} \sum_{\gamma, \beta} 
{\cal S}_{\alpha^{\prime}}[\gamma\bigl(\sigma(2), \ldots, \sigma(j-1)\bigr) | \sigma(2, \ldots, j-1) ]_{k_1}
\nn\\
&\times& 
{\cal S}_{\alpha^{\prime}}[\beta\bigl(\sigma(j), \ldots, \sigma(N-2)\bigr) | \sigma(j, \ldots, N-2) ]_{k_{N-1}}
A_{N}(1,\sigma(2,\ldots,N-2),N-1,N) \bigg|_{\alpha^{\prime} \to \alpha^{\prime}/4} 
\nn\\
&\times& 
\tilde{A}_N \left( \gamma\bigl(\sigma(2), \ldots, \sigma(j-1)\bigr),1,N-1, 
\beta\bigl(\sigma(j), \ldots, \sigma(n-2)\bigr),N \right) \bigg|_{\alpha^{\prime} \to \alpha^{\prime}/4}
\eea
where we sum over all orderings, $\sigma(1,\ldots,N)$, $\gamma(1,\ldots,N)$ and $\beta(1,\ldots,N)$ denote permutations of the $N!$ labels and the gauge amplitudes $A, \tilde{A}$ are associated with the left-moving and right-moving sectors, respectively. The momentum kernel ${\cal S}_{\alpha^{\prime}}$ is defined in Ref.~\cite{Bjerrum-Bohr:2010pnr}, but we quote its expression in the appendix. The amplitude can also be written as
\bea
M_N &=& (-1)^{N-3} \sum_{\sigma,\gamma} 
{\cal S}_{\alpha^{\prime}}[\gamma(2,\ldots,N-2) | \sigma(2,\ldots,N-2)]_{k_1}
\nn\\
&\times&
A_N(1,\sigma(2,\ldots,N-2),N-1,N) \bigg|_{\alpha^{\prime} \to \alpha^{\prime}/4} 
\tilde{A}_N(N-1,N,\gamma(2,\ldots,N-2),1) \bigg|_{\alpha^{\prime} \to \alpha^{\prime}/4} .
\eea
Recall that, on the leading Regge trajectory for the closed string, the spin is $S = 2 (n + 1)$, where $m^2 \alpha^{\prime} = 4 n$ for the mass level $n$.

For the Compton amplitude, the KLT relation reads
\bea
M_{4}({\bf 1}^S, 2^{h_2 \tilde{h}_2}, {\bf 3}^S,4^{h_4 \tilde{h}_4})
&=& (-1)^{n} \left( \frac{\pi \alpha^{\prime}}{4} \right)^{-1}
\sin\left( \frac{\pi \alpha^{\prime} s}{4} \right)
\nn\\
&\times& 
A_4({\bf 1}^{S/2},2^{h_2},{\bf 3}^{S/2},4^{h_4}) \bigg|_{\alpha^{\prime} \to \alpha^{\prime}/4} 
\tilde{A}_4({\bf 1}^{S/2},2^{\tilde{h}_2},4^{\tilde{h}_4},{\bf 3}^{S/2}) \bigg|_{\alpha^{\prime} \to \alpha^{\prime}/4}
\eea
where we used reflection symmetry for the right-moving sector and the gauge amplitudes are color-ordered amplitudes, see appendix for definitions. In order to describe interactions with gravitons, we have to correlate the helicities in the gauge amplitudes, $h_2 = \tilde{h}_2$ and $h_4 = \tilde{h}_4$. Notice that we are also correlating the spins of the massive particles. In our previous notation, we can write the amplitude as
\bea
M^{\alpha\gamma,\beta\delta}_{4}({\bf 1}, 2, {\bf 3},4) &=&
(-1)^{n} \left( \frac{\pi \alpha^{\prime}}{4} \right)^{-1}
\sin\left( \frac{\pi \alpha^{\prime} s}{4} \right) \sin\left( \frac{\pi \alpha^{\prime} u}{4} \right)
\nn\\
&\times& \left\{ \frac{{\cal V}_t}{\sin(\alpha^{\prime} \pi t) \alpha^{\prime} t \, \prod_{k=1}^{2n-1} (\alpha^{\prime} t - k)}
\frac{{\cal V}_t}{\alpha^{\prime} t \, \prod_{l=1}^{2n-1} (\alpha^{\prime} t - l)}
{\cal A}_0^{\alpha\beta} {\cal A}_0^{\gamma\delta} \right\}_{\alpha^{\prime} \to \alpha^{\prime}/4} 
\eea
where we used the cyclic property for the gauge color-ordered amplitude and monodromy relations. As before higher-spin indices are left implicit. The above open-string amplitudes were stripped from the factor $2 i e^2$.

\section{Electromagnetic Compton amplitudes for all spins}

\subsection{Soft expansion}

Our aim is to extract from the above string results terms potentially relevant for the classical description of spinning compact binary systems. Superstring amplitudes are famously defined in ten dimensions, so we need to perform a compactification down to four dimensions; we follow the prescription given by Ref.~\cite{Cangemi:2022abk} and define suitable four-dimensional polarizations and momenta. Moreover, we also isolate the poles corresponding to the mass level $n$; these are connected to the s-channel and the u-channel (since we consider electromagnetic amplitudes, the t-channel is absent). To extract our candidate field-theory amplitudes for generic spins, we begin by considering both momenta associated with the external photons as being soft, $q \to \hbar q$. To follow this procedure, we must conveniently rewrite the amplitude by using that
\bea
\frac{\Gamma (1-s \alpha^{\prime})\Gamma (1-u \alpha^{\prime})}
{\Gamma(1+\alpha^{\prime} t - 2n)}
\frac{1}{\alpha^{\prime} t \, \prod_{k=1}^{2n-1} (\alpha^{\prime} t - k)}
&=& \frac{\Gamma (1+ 2 \alpha^{\prime} \, k_1 \cdot q_2 )}{(1-s \alpha^{\prime}) (2-s \alpha^{\prime}) \cdots 
(n-s \alpha^{\prime})}
\nn\\
&\times& \frac{\Gamma (1 + 2 \alpha^{\prime} \, k_1 \cdot q_4)}{(1-u \alpha^{\prime}) (2-u \alpha^{\prime}) \cdots 
(n-u \alpha^{\prime})} 
\nn\\
&\times& \frac{1}{ \Gamma(1+\alpha^{\prime} t)}
\label{10}
\eea
where the recurrence formula
$$
\Gamma(z) = \frac{\Gamma(z+n+1)}{z (z+1) \cdots (z+n)} 
$$
as well as the result
$$
\Gamma(z-n) = (-1)^{n-1} \frac{\Gamma(-z) \Gamma(1+z)}{\Gamma(n+1-z)}
$$
were used. Observe that keeping only the level $n$ poles corresponds to the leading $1/\hbar$ term in the soft expansion. Then at leading order in a soft expansion we find
\beq
A^{\alpha\beta}_{t} \Bigg|_{\textrm{field theory}} \equiv {\cal A}^{\alpha\beta}_{S}
= \frac{1}{[\Gamma(n)]^2}
\frac{{\cal A}_0^{\alpha\beta}}{( s \alpha^{\prime} - n ) ( u \alpha^{\prime} -n )} .
\eeq
where the subscript $S$ in ${\cal A}^{\alpha\beta}_{S}$ denotes the spin of the massive particle. 

\subsection{First mass level}

Before continuing the discussion for generic spins, let us study the lowest-spin massive boson. The polarization tensors are given by $\epsilon_{a_1 a_S}({\bf 1}) = \epsilon_{a_1}({\bf 1}) \epsilon_{a_S}({\bf 1})$ and 
$\epsilon_{c_1 c_S}({\bf 3}) = \epsilon_{c_1}({\bf 3}) \epsilon_{c_S}({\bf 3})$. By taking a close look at the corresponding expression, one quickly finds terms that describe the Compton amplitude for scattering of massive spin $1$ particles. To separate such terms from the rest we consider a suitable compactification of the polarization vectors $\epsilon_{a_1}({\bf 1})$ and $\epsilon_{c_1}({\bf 3})$ -- we take these to be in the ``fifth dimension'' (such as $\epsilon^{d=10}({\bf 1}) = (0,0,0,0,1,{\bf 0})$). After performing this compactification, contraction with the polarizations, and a few algebraic manipulations, we arrive at the following expression of the Compton amplitude describing the interaction between photons and massive spin-$1$ particles:
\bea
{\cal A}_{1}({\bf 1},2^{h_2}, {\bf 3}, 4^{h_4}) &=& - \frac{2 i e^2}{(s-m^2) (u-m^2)}
\biggl\{ 2 \epsilon({\bf 1}) \cdot \epsilon({\bf 3}) \, k_1 \cdot f_2 \cdot f_4 \cdot k_1 
\nn\\
&+&  \left[ k_1 \cdot f_2 \cdot k_3 \, f_4^{\alpha\beta} 
+ k_1 \cdot f_4 \cdot k_3 \, f_2^{\alpha\beta} 
+ \frac12 k_1 \cdot ( q_4 - q_2 ) \bigl( f_2^{\alpha\rho} f_{4 \rho}^{\ \ \beta}
- f_2^{\beta\rho} f_{4 \rho}^{\ \ \alpha} \bigr) \right] {\cal J}_{\alpha\beta}
\nn\\
&-& \frac{q_2 \cdot q_4}{8} \bigl( f_{2 \alpha\beta} f_{4 \lambda\kappa}
+ f_{4 \alpha\beta} f_{2 \lambda\kappa} \bigr)  \{ {\cal J}^{\alpha\beta}, {\cal J}^{\lambda\kappa} \}
\label{14}
\biggr\}
\eea
where $h_2, h_4$ are the helicities of the photons and $f_{i}^{\mu\nu} = 2 q_i^{[\mu} \epsilon^{\nu]}(i) $. To obtain this expression we wrote $\epsilon_{a_S}({\bf 1}) \epsilon_{c_S}({\bf 3})
= {\cal J}_{a_S c_S}/2 + \epsilon_{(a_S}({\bf 1}) \epsilon_{c_S)}({\bf 3})$, where we have defined ${\cal J}^{\alpha\gamma} \equiv - i \epsilon^{\mu}({\bf 1}) ( M^{\alpha\gamma} )_{\mu\nu} \epsilon^{\nu}({\bf 3})$, $(M^{ab})_{c}^{\ d} = 2 i \delta^{[a}_{\ c} \eta^{b] d}$ being the Lorentz generator in the spin-$1$ representation. Furthermore:
\beq
\frac12 \{ {\cal J}^{\alpha\beta}, {\cal J}^{\lambda\kappa} \} \equiv
- \epsilon^{\mu}({\bf 1}) \bigl[ ( M^{\alpha\beta} )_{\mu\rho} ( M^{\lambda\kappa} )^{\rho}_{\ \nu} 
+ ( M^{\lambda\kappa} )_{\mu\rho} ( M^{\alpha\beta} )^{\rho}_{\ \nu}  \bigr] \epsilon^{\nu}({\bf 3}).
\eeq
Up to a numerical overall factor, the result~(\ref{14}) perfectly matches the one previously derived by others~\cite{Bautista:2022wjf} (except for possible sign differences due to different conventions on momenta). 

The spin multipole decomposition of the Compton amplitude given by~(\ref{14}) can be easily rewritten in the massive spinor-helicity basis~\cite{Bautista:2019tdr}. For the opposite helicity amplitude, 
$h_2 = +, h_4 = -$, one finds that
\bea
A_{S}({\bf 1},2^{+}, {\bf 3}, 4^{-}) &=& - \frac{2 i e^2}{m^{2S}} \frac{\langle 4 | {\bf 1} | 2 \bigr]^{2}}{(s-m^2) (u-m^2)}
\langle {\bf 3} |^{2S} \exp\left( i \frac{q_4^{\mu} \bar{\epsilon}_{4}^{\nu} {\cal J}_{1 \mu\nu}}
{\bar{p} \cdot \bar{\epsilon}_{4}} \right)  | {\bf 1} \rangle^{2S}
\nn\\
&=&  - \frac{2 i e^2}{(s-m^2) (u-m^2)}
\langle 4| {\bf 1} |2 \bigl]^{2-2S}
\Bigl( \langle {\bf 3} 4 \rangle \bigl[ {\bf 1} 2 \bigr] + \langle {\bf 1} 4 \rangle \bigl[ {\bf 3} 2 \bigr] \Bigr)^{2S}
\,\,\,\,\,\,\,\,
(S \leq 1)
\eea
where $\bar{p} = (p_1 - p_3)/2$ and $\bar{\epsilon}_{4}$ is the polarization vector for $q_4$ with reference momentum fixed to be $q_2$, see also~\cite{Guevara:19a}. A similar calculation can be performed for the same-helicity amplitude. Therefore the spin-1 amplitude, computed from a string-theory based calculation, agrees with other known expressions in the literature~\cite{Arkani-Hamed:2017jhn,Johansson:19}.

Observe from Eq.~(\ref{10}) that $\alpha^{\prime}$ is effectively parameterizing the expansion in the soft photon momenta, since it always appears accompanied by a Mandelstam variable. Nevertheless, to see that the approach of taking directly $\alpha^{\prime} \to 0$ is not, in general, the proper way to obtain the field theory limit which is useful in the present context, let us quote the amplitude for the spin-2 leading Regge state with this prescription. The calculation is rather involved but straightforward. After the dust settles, we find that
\bea
{\cal A}_{2}({\bf 1},2^{h_2}, {\bf 3}, 4^{h_4}) &=& - \frac{2 i e^2}{\left( s-m^2 \right)\left( u-m^2 \right)}
\Bigg\{ 2 \epsilon({\bf 1}) \cdot \epsilon({\bf 3}) k_1 \cdot f_2 \cdot f_4 \cdot k_1
\nn\\
&+& 2 \Bigg[ k_1 \cdot f_2 \cdot k_3 f_4^{\mu \nu} + k_1 \cdot f_4 \cdot k_3 f_2^{\mu \nu}
+ \frac{1}{2} k_1 \cdot \left(q_4 - q_2\right) 
\left( f_2^{\mu \rho} f_{4 \rho}^{\ \ \nu} - f_2^{\nu \rho} f_{4 \rho}^{\ \ \mu} \right) \Bigg] {\cal J}_{\mu \nu}
\nn\\
&+& \frac{q_2 \cdot q_4}{2} \Big(f_{2 \mu \nu} f_{4 \rho \sigma} + f_{2 \rho \sigma} f_{4 \mu \nu} \Big)
\left( \frac12 \{ {\cal J}^{\mu\nu}, {\cal J}^{\rho\sigma} \} 
+ 8 \epsilon^{\mu \sigma}({\bf 1}) \epsilon^{\nu \rho}({\bf 3}) \right) \Bigg\}
\label{JLtriste}
\eea
where ${\cal J}^{ab} \equiv - \frac{i}{2} \epsilon^{c_1c_2}({\bf 1}) {(M^{a b})_{c_1c_2}}^{d_1d_2}  
\epsilon_{d_1d_2}({\bf 3})$, ${(M^{a b})_{c(2)}}^{d(2)} = 4 i \delta_{(c_1}^{[a} \eta^{b](d_1} 
\delta_{c_2)}^{d_2)}$ being the Lorentz generator in the spin-$2$ representation. This amplitude in the classical limit will produce contributions up to the quadrupole term. However, for a spin-$S$ quantum particle, the classical counterpart should produce an expansion up to the $2S$-th power of the classical spin. Besides, strictly speaking, the $\alpha^{\prime} \to 0$ amounts to projecting out all massive spinning particles from the spectrum. These are clearly indications that the straightforward limit $\alpha^{\prime} \to 0$ is not adequate to our purposes.

\subsection{All mass levels}

Given the precedent successful computation of the amplitude for the massive spin-$1$ case, Eq.~(\ref{14}), we now undertake the task of investigating the general case, with the polarization tensors given by
$\epsilon_{a_1 \cdots a_S}({\bf 1}) = \epsilon_{a_1}({\bf 1}) \ldots \epsilon_{a_n}({\bf 1}) 
\epsilon_{a_S}({\bf 1})$ and 
$\epsilon_{c_1 \cdots c_S}({\bf 3}) = \epsilon_{c_1}({\bf 3}) \ldots \epsilon_{c_n}({\bf 3}) 
\epsilon_{c_S}({\bf 3})$. Following previous prescriptions, we consider the compactification of two massive polarization vectors, namely $\epsilon_{a_n}({\bf 1})$ and $\epsilon_{c_n}({\bf 3})$. Furthermore, we consider only terms with $k=l$ in the double summations in the full expression for  ${\cal A}_0^{\alpha\beta}$ (see appendix) as these are the ones that have explicit inverse propagators in the numerators. Our candidate electromagnetic Compton amplitude takes the form
\bea
{\cal A}_S^{\alpha\beta} &=&
\frac{2 i e^2 }{\left(s-m^2\right) \left(u-m^2\right)} 
\sum_{j=0}^{n-1} \sum_{k=0}^{n-1-j}
\frac{(-1)^{n-1-j} n^2 }{\Gamma (j+2) \Gamma (k+1)^2 \Gamma (-j-k+n)^2}
\nn\\
&\times& 
(2 \alpha^{\prime})^{n-1-j} 
\eta^{a(j) c(j)}
\gamma(s \alpha^{\prime},n-1) \gamma(u \alpha^{\prime},n-1)
q_2^{c(n-1-j-k)} q_4^{a(k)} q_2^{a(n-1-j-k)} q_4^{c(k)}
\nn\\
&\times& \Biggl\{  - (u  - m^2) \left[ 
\Bigl( -  \eta^{a_s c_s} k_3^{\beta} + \eta^{a_s \beta} q_4^{c_s} - \eta^{c_s \beta} q_4^{a_s}  \Bigr) \Bigl( k_3^{\alpha} + q_4^{\alpha} \Bigr) 
+ q_2^{a_s} \left(  \eta^{\alpha \beta} q_4^{c_s} - \eta^{c_s \beta} q_4^{\alpha}  \right)
\right.
\nn\\
&+& \left. \eta^{a_s \alpha} \Bigl( (q_2 \cdot q_4) \eta^{c_s \beta} - q_2^{\beta} q_4^{c_s} \Bigr)
+ k_3^{\beta} \left( q_2^{c_s} \eta^{a_s \alpha} - q_2^{a_s} \eta^{\alpha c_s} \right)
 \right]
\nn\\
&-& (s - m^2) \left[ 
\Bigl( - \eta^{a_s c_s} k_3^{\alpha} + q_2^{c_s} \eta^{a_s \alpha} - q_2^{a_s} \eta^{\alpha c_s}  \Bigr)
\Bigl(  k_3^{\beta} + q_2^{\beta} \Bigr) 
+ q_4^{a_s} \left( \eta^{\alpha\beta} q_2^{c_s} - \eta^{\alpha c_s} q_2^{\beta} \right)
\right.
\nn\\
&+& \left. \eta^{a_s \beta} \Bigl( (q_2 \cdot q_4) \eta^{\alpha c_s}  - q_2^{c_s} q_4^{\alpha}  \Bigr) 
+ k_3^{\alpha} \left( q_4^{c_s} \eta^{a_s \beta} - q_4^{a_s} \eta^{c_s \beta}  \right)
\right]
\nn\\
&+& \frac{1}{2} \eta^{\alpha\beta} \eta^{a_s c_s} (s-m^2) (u-m^2) 
\Biggr\},
\label{26}
\eea
where $\gamma(x,N) \equiv \prod_{m=1}^{N} (-x + m) = (1-x)_N$. This amplitude is gauge invariant in the sense that $ q_{4, \beta} {\cal A}^{\alpha\beta}_n = 0 = q_{2, \alpha} 
{\cal A}^{\alpha\beta}_n$. The number of the mass level $n$ characterizes the spin of the massive particle under consideration and we can just take $n \to S$ in the above expressions. Moreover, we remark that this prescription is valid to all mass levels $n \geq 1$.

The Compton amplitude given by Eq.~(\ref{26}) is the full quantum-field-theory amplitude after imposing the two aforementioned constraints, that is, messengers with soft momenta and adequate compactification of one of the massive polarization vectors. As we can see, there are many contributions to the full amplitude, many of them are unlikely to play a role in the root-Kerr problem. However, we will now make the following conjecture: The amplitude whose classical limit describes the dynamics of a compact body is secretly hidden in the full quantum amplitude~(\ref{26}) as a specific contribution. To expose such a contribution, it is obvious that the two requirements established above are not enough; instead, there should be at least one extra step.

We now require the associated residues to be given by the product of two {\it minimally coupled} 3-point amplitudes. For this purpose, let us analyze the s-channel cut (similar considerations apply to the u-channel cut). We take particle 3 to be a small boost of particle 1 (or the inverse)~\cite{Cangemi:2022abk}:
\bea
|3^a \rangle &=& |1^a \rangle + \frac{m}{2 m_q^2} \, \slashed{q} |1^a \bigl]
\nn\\
|3^a \bigl] &=& - |1^a \bigl] - \frac{m}{2 m_q^2} \, \slashed{q} |1^a \rangle,
\eea
which implies that
\bea
\frac{1}{m} \langle {\bf 1} {\bf 3} \rangle &=& \mathbbm{1} + \frac{1}{2 m_q^2} \, \langle {\bf 1} | q | {\bf 1} \bigl]
\nn\\
\frac{1}{m} \bigl[ {\bf 1} {\bf 3} \bigr] &=& \mathbbm{1} - \frac{1}{2 m_q^2} \, \langle {\bf 1} | q | {\bf 1} \bigl]
\nn\\
\epsilon^\mu(\mathbf{3}) &=& \bar\epsilon^\mu(\mathbf{1}) 
- \frac{q\cdot\bar\epsilon(\mathbf{1})}{m_q^2}\left(k_1^\mu + \frac{1}{2}q^\mu\right)
\label{31}
\eea
where $m_q = m \sqrt{1+ q^2/4m^2}$ and $q \equiv -k_3 -k_1 = q_2 + q_4$ is the momentum transferred, which is very small compared to the massive momenta. Therefore we can write that
\bea
( q_2 \cdot \epsilon({\bf 3}) \, q_2 \cdot \epsilon({\bf 1}) )^l
&\approx&
( q_2 \cdot \bar\epsilon(\mathbf{1}) \, q_2 \cdot \epsilon({\bf 1}) )^l
\nn\\
( q_4 \cdot \epsilon({\bf 3}) \, q_4 \cdot \epsilon({\bf 1}) )^l
&\approx&
( q_4 \cdot \epsilon({\bf 3}) \, q_4 \cdot \bar\epsilon(\mathbf{3}) )^l
\nn\\
\bigl( \epsilon(\mathbf{3}) \cdot \epsilon(\mathbf{1}) \bigr)^j &=& 
\left( \frac{2}{m} \langle {\bf 3} {\bf 1} \rangle \right)^j
\sum_{p=0}^{j} \frac{j!}{p!(j-p)!} \left( \frac{1}{2 m} \langle {\bf 3} {\bf 1} \rangle \right)^p .
\eea
The s-channel residue reads (after contracting with external polarizations)
\bea
{\cal A}_n\Bigg|_{\textrm{s-channel residue}} &\approx& - 2 i e^2
\sum_{j=0}^{n-1} \sum_{k=0}^{n-1-j} \sum_{p=0}^{j} 
\frac{j!}{p!(j-p)!}
\frac{(-1)^{n-1-j} n^2 (2 n)^{n-1-j} [\gamma(n,n-1)]^2 }
{\Gamma (j+2) \Gamma (k+1)^2 \Gamma (-j-k+n)^2}
\nn\\
&\times& 
\left( \frac{2}{m} \langle {\bf 3} {\bf 1} \rangle \right)^j
 \left( \frac{1}{2 m} \langle {\bf 3} {\bf 1} \rangle \right)^p
\left( \frac{\bigl( q_2 \cdot \bar{\epsilon}({\bf 1}) \, q_2 \cdot \epsilon({\bf 1}) \bigr)}{m^2} \right)^{n-1-j-k} 
\left( \frac{\bigl( q_4 \cdot \bar{\epsilon}({\bf 3}) \, q_4 \cdot \epsilon({\bf 3}) \bigr)}{m^2} \right)^{k} 
\nn\\
&\times& 
\epsilon_{a_s}({\bf 1}) \Big[ 
\Bigl( -  \eta^{a_s c_s} k_3 \cdot \epsilon(4) + \epsilon^{a_s}(4) q_4^{c_s} 
- \epsilon^{c_s}(4) q_4^{a_s}  \Bigr) 
( k_3 + q_4 ) \cdot \epsilon(2)
\nn\\
&+& q_2^{a_s} \Big(  \epsilon(2) \cdot \epsilon(4) \, q_4^{c_s} 
- \epsilon^{c_s}(4) \, q_4 \cdot \epsilon(2) \Big)
\nn\\
&+& \epsilon^{a_s}(2) \Bigl( (q_2 \cdot q_4) \, \epsilon^{c_s}(4) 
- q_2 \cdot \epsilon(4) \, q_4^{c_s} \Bigr)
\nn\\
&+& ( k_3 \cdot \epsilon(4) ) \Big( q_2^{c_s} \epsilon^{a_s}(2) - q_2^{a_s} \epsilon^{c_s}(2) \Big)
\Big ] \epsilon_{c_s}({\bf 3}) .
\label{27}
\eea
where, after using s-channel kinematics and the fact that $q^2 \lll 1/\alpha^{\prime}$, we find that
$\gamma(s \alpha^{\prime},n-1) = \gamma(n,n-1)$ and 
$\gamma(u \alpha^{\prime},n-1) = \gamma(n - t \alpha^{\prime},n-1) \approx \gamma(n,n-1)$ at leading order. To understand how we should impose consistent factorization as a constraint, let us analyze the corresponding ``helicity amplitudes" separately. First let us consider the all plus amplitude. The term in square brackets is just the spin multipole decomposition of the QED Compton amplitude for matter of spin one~\cite{Bautista:2022wjf}, so this must be proportional to $\bigl[ 24 \bigr]^2 \langle {\bf 1} {\bf 3} \rangle^{2}/t$ in the spinor-helicity basis~\cite{Johansson:19}. So the rest of the amplitude should produce something like $m^{2-2S} \langle {\bf 1} {\bf 3} \rangle^{2S-2}$. The only term that furnishes this is given by the contribution $p = j = n-1$:
\bea
\bar{\cal A}_n(++)\Bigg|_{\textrm{s-channel residue}} &=& 2 i e^2 m^{2-2n}
 \bigl[ 24 \bigr]^2
\langle {\bf 1} {\bf 3} \rangle^{2n}
\frac{1}{u - m^2} 
\left(\frac{ n^2  [\gamma(n,n-1)]^2 }{\Gamma (n+1)} \right)
\eea
where we have denoted by a bar such a contribution. We also used s-channel kinematics. Apart from the positive numerical term inside the brackets, this term reproduces the correct s-channel factorization. Now let us focus on the amplitude with opposite helicities. As the term in square brackets in Eq.~(\ref{27}) is roughly the QED Compton amplitude for matter of spin one, in the spinor-helicity basis this should be proportional to
$$
\frac{\langle 4 | {\bf 1} | 2 \bigr]^2}{ m^4 t} \big| \langle {\bf 1} | P | {\bf 3} \bigr] \big|^2 
$$
where $P = - k_1 - q_2 = k_3 + q_4$. However, since $| {\bf 1} \rangle$ and $| \bar{\bf 1} \rangle$ are related by complex conjugation (the same for ${\bf 3}$), we find that
\bea
q_2 \cdot \bar{\epsilon}({\bf 1}) \, q_2 \cdot \epsilon({\bf 1}) &=& 
- \frac{ | \langle {\bf 1} | P | {\bf 1} \bigr] |^2  }{2 m^2}
\approx - \frac{ | \langle {\bf 1} | P | {\bf 3} \bigr] |^2  }{2 m^2}
\nn\\
q_4 \cdot \bar{\epsilon}({\bf 3}) \, q_4 \cdot \epsilon({\bf 3}) &=&
- \frac{ | \langle {\bf 3} | P | {\bf 3} \bigr] |^2  }{2 m^2}
\approx - \frac{ | \langle {\bf 1} | P | {\bf 3} \bigr] |^2  }{2 m^2},
\eea
up to subleading terms in the classical limit, where we used that 
$k_1 \cdot \epsilon({\bf 1}) = k_1 \cdot \bar{\epsilon}({\bf 1}) = 0$, and similarly for ${\bf 3}$. Only the term $j=0$ reproduces the proper s-channel residue; hence
\beq
\bar{\cal A}_n(+-)\Bigg|_{\textrm{s-channel residue}} \approx 2 i e^2
\frac{ \langle 4 | {\bf 1} | 2 \bigr]^2 }{ m^{4n} (u-m^2)} \big| \langle {\bf 1} | P | {\bf 3} \bigr] \big|^{2n} 
\left(  \frac{4^{n-1} n^{n+1} \Gamma \left(n-\frac{1}{2}\right) [\gamma(n,n-1)]^2}
{\sqrt{\pi } \Gamma (n)^3} \right)
\eeq
where, with abuse of notation, we again denoted by a bar this specific term. We also used the result:
$$
\sum_{k=0}^{n-1} \frac{1}{ \Gamma (k+1)^2 \Gamma (-k+n)^2} =
\frac{4^{n-1} \Gamma \left(n-\frac{1}{2}\right)}{\sqrt{\pi } \Gamma (n)^3} .
$$
Again apart from the positive numerical term inside the brackets, this term reproduces the correct s-channel factorization for the amplitude with opposite helicities.

Let us state our proposal for the electromagnetic Compton scattering amplitude. Consider first the amplitude with opposite helicities. Denoting by a bar this particular contribution we find, for a massive particle with spin $s=n>0$ integer:
\beq
\bar{{\cal A}}_{n}({\bf 1},2^{+}, {\bf 3}, 4^{-}) =  \frac{2 i e^2 }{\left(s-m^2\right) \left(u-m^2\right)} 
\mathbb{A}_{n}({\bf 1}, 2^{+}, {\bf 3}, 4^{-})
\label{42b}
\eeq
where
\bea
\mathbb{A}_{n}({\bf 1}, 2^{+}, {\bf 3}, 4^{-}) &\equiv& 
 {\cal N} \frac{n^2 (2 n)^{n-1} (-1)^{n-1}}{\Gamma (n)^2}
\gamma(s \alpha^{\prime},n-1) \gamma(u \alpha^{\prime},n-1)
\nn\\
&\times&
\left( \frac{\bigl( q_2 \cdot \epsilon({\bf 3}) \, q_2 \cdot \epsilon({\bf 1}) \bigr)}{m^2} \right)^{n-1} 
  \, _2F_1\left(1-n,1-n;1; \frac{\bigl( q_4 \cdot \epsilon({\bf 1}) \, q_4 \cdot \epsilon({\bf 3}) \bigr)}
 {\bigl( q_2 \cdot \epsilon({\bf 3}) \, q_2 \cdot \epsilon({\bf 1}) \bigr)}\right)
\nn\\
&\times& \epsilon^{+}_{\alpha}(2) \epsilon^{-}_{\beta}(4)
\Biggl\{  - (u  - m^2) \left[ 
\Bigl( -  \eta^{a_s c_s} k_3^{\beta} + \eta^{a_s \beta} q_4^{c_s} - \eta^{c_s \beta} q_4^{a_s}  \Bigr) \Bigl( k_3^{\alpha} + q_4^{\alpha} \Bigr) 
\right.
\nn\\
&+& \left. q_2^{a_s} \left(  \eta^{\alpha \beta} q_4^{c_s} - \eta^{c_s \beta} q_4^{\alpha}  \right)
\right.
\nn\\
&+& \left. \eta^{a_s \alpha} \Bigl( (q_2 \cdot q_4) \eta^{c_s \beta} - q_2^{\beta} q_4^{c_s} \Bigr)
+ k_3^{\beta} \left( q_2^{c_s} \eta^{a_s \alpha} - q_2^{a_s} \eta^{\alpha c_s} \right)
 \right]
\nn\\
&-& (s - m^2) \left[ 
\Bigl( - \eta^{a_s c_s} k_3^{\alpha} + q_2^{c_s} \eta^{a_s \alpha} - q_2^{a_s} \eta^{\alpha c_s}  \Bigr)
\Bigl(  k_3^{\beta} + q_2^{\beta} \Bigr) 
+ q_4^{a_s} \left( \eta^{\alpha\beta} q_2^{c_s} - \eta^{\alpha c_s} q_2^{\beta} \right)
\right.
\nn\\
&+& \left. \eta^{a_s \beta} \Bigl( (q_2 \cdot q_4) \eta^{\alpha c_s}  - q_2^{c_s} q_4^{\alpha}  \Bigr) 
+ k_3^{\alpha} \left( q_4^{c_s} \eta^{a_s \beta} - q_4^{a_s} \eta^{c_s \beta}  \right)
\right]
\nn\\
&+& \frac{1}{2} \eta^{\alpha\beta} \eta^{a_s c_s} (s-m^2) (u-m^2) 
\Biggr\} \epsilon_{a_s}({\bf 1}) \epsilon_{c_s}({\bf 3}) 
\label{42}
\eea
where we used that
$$
\sum _{k=0}^{n-1} \frac{B^k A^{-k+n-1}}{\Gamma (k+1)^2 \Gamma (n-k)^2}
= \frac{A^{n-1} \, _2F_1\left(1-n,1-n;1;\frac{B}{A}\right)}{\Gamma (n)^2}
$$
and $\, _2F_1\left(a,b;c;x\right)$ is the hypergeometric function. Notice that we have added an arbitrary normalization factor ${\cal N}(n)$ (with ${\cal N}(1) = 1$) in order to take care of the multiplicative numerical factor that appears in the factorization of the physical channels. That is,
$$
{\cal N}(n) = \left(  \frac{4^{n-1} n^{n+1} \Gamma \left(n-\frac{1}{2}\right) [\gamma(n,n-1)]^2}
{\sqrt{\pi } \Gamma (n)^3} \right)^{-1}.
$$
As for the all plus amplitude, as discussed we obtain the usual result (up to an irrelevant positive constant) which we quote here for completeness:
\beq
\bar{{\cal A}}_{n}({\bf 1},2^{+}, {\bf 3}, 4^{+}) =  2 i e^2 m^{2-2n}
 \frac{ \bigl[ 24 \bigr]^2 \langle {\bf 1} {\bf 3} \rangle^{2n} }{(s-m^2)(u - m^2)} 
\left(\frac{ n^2  [\gamma(n,n-1)]^2 }{\Gamma (n+1)} \right) .
\eeq
As in the opposite-helicity case, the positive numerical factor found above can be canceled by introducing an arbitrary suitable normalization factor. Finally, the other helicity amplitudes can be obtained from the above ones by resorting to the usual bracket swap, $\langle (\cdots) \rangle \leftrightarrow [ (\cdots) ]$, in a suitable spinor-helicity basis.

In summary, our claim is: Open string-theory amplitudes can potentially contain amplitudes that describe compact astrophysical bodies as specific contributions (among many others) in disguise inside the full amplitude~\footnote{Indeed, as argued in Ref.~\cite{Chung:19}, 3-point string amplitudes produce couplings which are ``maximally complex'' in that it represents the most general form of the three-point amplitude for one massless and two equal mass legs with spin $S$ but with the corresponding couplings assuming definite values. This amplitude includes, of course, the minimal coupling amplitude which describes root-Kerr in the classical limit.}. To extract this hidden contribution to the amplitude, all other contributions must be projected out, and we propose that this can be achieved by imposing the following three constraints:

\begin{itemize}

\item Soft momentum transfer, $q^2 \lll 1/\alpha^{\prime}$; 

\item Compactification of the polarizations $\epsilon^{a_n}({\bf 1})$ and $\epsilon^{c_n}({\bf 3})$, so that 
$n \to S$; 

\item Consistent factorization in the physical channels for any helicity choice.

\end{itemize}

Let us see how such a reasoning can also be used to extract the correct 3-point root-Kerr amplitude from the corresponding superstring amplitude. The relevant three-point amplitude is given by~\cite{Schlotterer:2010kk,Cangemi:2022abk}
\bea
A_3({\bf 1}, {\bf 2}, q) &=& e (2 \alpha^{\prime})^S (S-1)!
\sum_{j=0}^{S} \frac{(-\epsilon({\bf 1}) \cdot \epsilon({\bf 2}) )^j}{(2 \alpha^{\prime})^j j! [(S-j)!]^2}
\nn\\
&\times& \left( - j (\epsilon(q) \cdot k_1) ( - \epsilon({\bf 1}) \cdot q \, \epsilon({\bf 2}) \cdot q )^{S-j}
+ \frac{S(S-j)}{2 \alpha^{\prime}} \epsilon_{\mu}({\bf 2}) f^{\mu\nu} \epsilon_{\nu}({\bf 1})
( - \epsilon({\bf 1}) \cdot q \, \epsilon({\bf 2}) \cdot q )^{S-j-1}
\right)
\nn\\
\eea
where, as above, $f^{\mu\nu} = 2 q^{[\mu} \epsilon^{\nu]}(q)$.  The field-theory limit proposed here requires a soft expansion and, as argued before, this is parametrized by $\alpha^{\prime}$. Moreover, we also consider the compactification mentioned above for the massive polarizations. This effectively implies only the term $j=S$ in the above sum is kept, which leads us to
\bea
A_3({\bf 1}, {\bf 2}, q) = \frac{e}{\sqrt{2}} (-1)^{S} m x_1 ( \epsilon({\bf 1}) \cdot \epsilon({\bf 2}) )^{S}  
\eea
where the $x_1$ factor is defined as usual by $x_1 = \sqrt{2} ( k_1\cdot\epsilon(q) )/m$. In the classical limit, $x_1$ is $m$-independent~\cite{Guevara:19a}. On the other hand, we can use that
\beq
( \epsilon({\bf 1}) \cdot \epsilon({\bf 2}) )^S = \frac{1}{m^{2S}} 
\langle {\bf 2} {\bf 1} \rangle^S \bigl[ {\bf 1} {\bf 2} \bigr]^S
\eeq
and, from three-point special kinematics:
\beq
\bigl[ {\bf 1} {\bf 2} \bigr] = \langle {\bf 1} {\bf 2} \rangle 
- \frac{1}{m x_1} \langle {\bf 1} q \rangle \langle q {\bf 2} \rangle
\eeq
so we can argue that
\bea
A_3({\bf 1}, {\bf 2}, q^+) &=& \frac{e}{\sqrt{2}} m x_1^+ 
\left( \frac{\langle {\bf 1} {\bf 2} \rangle}{m} \right)^{2S}
\Bigl( 1 + {\cal O}(1/m) \Bigr)
\nn\\
A_3({\bf 1}, {\bf 2}, q^-) &=& \frac{e}{\sqrt{2}} m x_1^- 
\left( \frac{\bigl[ {\bf 1} {\bf 2} \bigr]}{m} \right)^{2S}
\Bigl( 1 + {\cal O}(1/m) \Bigr) .
\eea
In the end, we are able to extract from the full 3-point amplitude the minimal coupling quantum amplitudes which describe root-Kerr in the classical limit.

\subsection{Classical limit of the opposite-helicity amplitude}

Now let us study the classical limit of Eq.~(\ref{42b}) ($\hbar \to 0$, $S \to \infty$, $\hbar s$ fixed) as this is the helicity set up that contributes to the 2PM classical amplitude. As mentioned above, there are many works designed to tackle the problem of providing an ansatz for a local Compton amplitude. For practical terms, we choose here to compare our results with the ones derived in Ref.~\cite{Cangemi:2023ysz} as their presentation is more suitable for our analysis. Throughout this subsection, we will employ many of their definitions and manipulations to facilitate comparison. Many of the details which we will not display here for the sake of brevity and clarity can be found in such a reference and we refer the interested reader to it.

One may start by rewriting Eq.~(\ref{42}) as
\bea
\mathbb{A}_{n}({\bf 1}, 2^{+}, {\bf 3}, 4^{-}) &=& 
- {\cal N} \frac{n^2 (2 n)^{n-1} (-1)^{n-1}}{\Gamma (n)^2}
\gamma(s \alpha^{\prime},n-1) \gamma(u \alpha^{\prime},n-1)
\Bigl( \langle {\bf 3} 4 \rangle \bigl[ {\bf 1} 2 \bigr] + \langle {\bf 1} 4 \rangle \bigl[ {\bf 3} 2 \bigr] \Bigr)^{2}
\nn\\
&\times&
\left( \frac{\bigl( (q_2 \cdot \rho)^2 - ( q_2 \cdot \bar{\rho} )^2  \bigr)}{2 m^4} \right)^{n-1} 
  \, _2F_1\left(1-n,1-n;1; \frac{\bigl( (q_4 \cdot \rho)^2 - ( q_4 \cdot \bar{\rho} )^2  \bigr)}
 {\bigl( (q_2 \cdot \rho)^2 - ( q_2 \cdot \bar{\rho} )^2  \bigr)}\right)
\label{42a}
\eea
where we wrote the product of massive polarizations in terms of the following two complex vectors~\cite{Cangemi:2023ysz}
\bea
\rho^\mu &\equiv& \frac{1}{2} \Big( \bra{\bf{3}}\sigma^\mu|\bf{1}]
 + \bra{\bf{1}}\sigma^\mu|\bf{3}] \Big) \,,
 \nn\\
\bar\rho^\mu &\equiv& \frac{1}{2} \Big( \bra{\bf{3}}\sigma^\mu|\bf{1}]
 - \bra{\bf{1}}\sigma^\mu|\bf{3}] \Big) \,.
\eea
The amplitude should display exchange symmetry between massless particles, which is not manifest here, but it is definitely inherited from the full string amplitude, as one can easily see from the above results and the appendix -- this follows from the usual symmetry contained in the binomial theorem,
$$
(x+y)^n = \sum_{k=0}^n {n \choose k}x^{n-k}y^k = \sum_{k=0}^n {n \choose k}x^{k}y^{n-k} .
$$
We should combine this exchange symmetry with parity. Furthermore, by introducing the quantities~\cite{Cangemi:2023ysz}
\bea
W_{\pm} &\equiv& \frac{m}{2} \Bigl( \langle {\bf 1} {\bf 3} \rangle \pm [ {\bf 1} {\bf 3} ] \Bigr)
\nn\\
U &\equiv& \frac12 \Bigl(  \bra{{\bf 1}} 4 |{\bf 3}]  - \bra{{\bf 3}} 4 |{\bf 1}] \Bigr)
- m  [ {\bf 1} {\bf 3} ]
\nn\\
V &\equiv& \frac12 \Bigl(  \bra{{\bf 1}} 4 |{\bf 3}] + \bra{{\bf 3}} 4 |{\bf 1}] \Bigr)
\nn\\
\chi^{\mu} &\equiv& \langle 4| \sigma^{\mu}| 2]
\eea
we can rewrite the amplitude as
\bea
\mathbb{A}_{n}({\bf 1}, 2^{+}, {\bf 3}, 4^{-}) &=& 
- {\cal N} \frac{n^2 (2 n)^{n-1} (-1)^{n-1}}{2 \, \Gamma (n)^2}
\gamma(s \alpha^{\prime},n-1) \gamma(u \alpha^{\prime},n-1)
( \rho \cdot \chi)^{2}
\nn\\
&\times&
\left[ \left( \frac{ V^2 - \Bigl( 4 W_{-} - (U + m  [ {\bf 1} {\bf 3} ]) \Bigr)^2 }{2 m^4} \right)^{n-1} 
\right.
\nn\\
&\times& \left.
  \, _2F_1\left(1-n,1-n;1; \frac{\Bigl( V^2 
  - (U + m  [ {\bf 1} {\bf 3} ])^2 \Bigr)}
 {V^2 - \Bigl( 4 W_{-} - (U + m  [ {\bf 1} {\bf 3} ]) \Bigr)^2}\right)
\right.
 \nn\\
&+& \left. 
\left( \frac{ V^2 - \Bigl( 4 W_{-} + U + m  \langle {\bf 1} {\bf 3} \rangle \Bigr)^2 }{2 m^4} \right)^{n-1} 
\right.
\nn\\
&\times& \left.
  \, _2F_1\left(1-n,1-n;1; \frac{\Bigl( V^2 
  - (U + m  \langle {\bf 1} {\bf 3} \rangle)^2 \Bigr)}
 {V^2 - \Bigl( 4 W_{-} + U + m  \langle {\bf 1} {\bf 3} \rangle \Bigr)^2}\right)
\right]
\eea
To obtain the classical amplitude, we adopt the approach based on coherent spin states, see discussion in Refs.~\cite{Aoude:2021oqj,Cangemi:2023ysz}. The idea here is to consider the scattering of coherent spin states in place of the massive spin particles. The classical amplitude is the $\hbar \to 0$ limit of an infinite sum over diagonal finite-spin amplitudes
\beq
A_{\textrm{cl}}({\bf 1},2^{+}, {\bf 3}, 4^{-}) = \lim_{\hbar \to 0} e^{- |{\cal Z}|^2}
\sum_{S=0}^{\infty} \frac{1}{(2S)!} \bar{{\cal A}}_{S}({\bf 1},2^{+}, {\bf 3}, 4^{-}) 	
\eeq
where non-diagonal contributions are neglected and the $\hbar \to 0$ limit is taken after resumming the coherent states. A thorough discussion of the various $\hbar$ scalings of spin variables in the coherent-state approach can be found in Ref.~\cite{Cangemi:2023ysz}. In turn, $|{\cal Z}|^2$ is the squared norm of the physical SU(2) wavefunction for a massive particle, described by SU(2) little-group spinors ${\cal Z}_a$ and $\bar{ {\cal Z} }_a$ 
($i=1,3$, $a=1,2$)
\bea
\ket{{\bf i}} &=& \ket{i^a} {\cal Z}_{i a} \,, \qquad \quad
\ket{\bar {\bf i}} = \ket{i^a} \bar{ {\cal Z} }_{i a} \,, 
\nn\\
|{\bf i}] &=& |i^a] {\cal Z}_{i a} \,, \qquad\,\quad
|\bar {\bf i}] = |i^a] \bar{ {\cal Z} }_{i a} \,, 
\eea
where the unbarred spinors are incoming, and the barred spinors are outgoing. In this case 
$|{\cal Z}|^2 = {\cal Z}_a \bar{ {\cal Z} }^a$ for each spinning particle, where 
${\cal Z}^a = \epsilon^{ab} {\cal Z}_b$, $\epsilon^{ab}$ being the SU(2) Levi-Civita symbol.

When using quantum amplitudes to calculate the classical potential, we know that classical contributions arise only from the region where internal matter lines are close to on shell~\cite{Bern:2019crd}. Using this reasoning, one can argue that, in the limit $\hbar \to 0$, we should have that $\gamma(s \alpha^{\prime},n-1) \gamma(u \alpha^{\prime},n-1) \approx  [\gamma(n,n-1)]^2$. This result can also be obtained with the help of a soft-expansion argument. Moreover, summing over all possible spin states in our case means that the sum should start at $n=1$. Hence we claim that the classical amplitude is obtained with
\bea
A_{\textrm{cl}}({\bf 1},2^{+}, {\bf 3}, 4^{-}) &=& 
- \lim_{\hbar \to 0} 
\frac{i e^2 }{\left(s-m^2\right) \left(u-m^2\right)} 
e^{- |{\cal Z}|^2} ( \rho \cdot \chi)^{2}
\nn\\
&\times&
\sum_{n=1}^{\infty} \frac{1}{(2 (n-1))!} 
\left(  \frac{\sqrt{\pi } \Gamma(n)}{ 4^{n-1} \Gamma\left(n-\frac{1}{2}\right) } \right)
(-1)^{n-1}
\left[ \left( \frac{ V^2 - \Bigl( 4 W_{-} - (U + m  [ {\bf 1} {\bf 3} ]) \Bigr)^2 }{2 m^4} \right)^{n-1} 
\right.
\nn\\
&\times& \left.
  \, _2F_1\left(1-n,1-n;1; \frac{\Bigl( V^2 
  - (U + m  [ {\bf 1} {\bf 3} ])^2 \Bigr)}
 {V^2 - \Bigl( 4 W_{-} - (U + m  [ {\bf 1} {\bf 3} ]) \Bigr)^2}\right)
\right.
 \nn\\
&+& \left. 
\left( \frac{ V^2 - \Bigl( 4 W_{-} + U + m  \langle {\bf 1} {\bf 3} \rangle \Bigr)^2 }{2 m^4} \right)^{n-1} 
\right.
\nn\\
&\times& \left.
  \, _2F_1\left(1-n,1-n;1; \frac{\Bigl( V^2 
  - (U + m  \langle {\bf 1} {\bf 3} \rangle)^2 \Bigr)}
 {V^2 - \Bigl( 4 W_{-} + U + m  \langle {\bf 1} {\bf 3} \rangle \Bigr)^2}\right)
\right] .
\eea
To deal with the term $e^{- |{\cal Z}|^2} ( \rho \cdot \chi)^{2}$, let us come back to the well established   expression (\ref{14}), which is valid for $S \leq 1$; its classical limit is given by
\bea
A_{\textrm{cl}}({\bf 1},2^{+}, {\bf 3}, 4^{-}) &=& - \lim_{\hbar \to 0} 
\frac{i e^2 \, (k_1 \cdot \chi)^2}{(s-m^2) (u-m^2)}
e^{- |{\cal Z}|^2}
\sum_{S=0}^{\infty} \frac{1}{(2S)!}
\left( \frac{\rho \cdot \chi}{k_1 \cdot \chi} \right)^{2S} + {\cal O}(a^3)
\nn\\
&=& - \lim_{\hbar \to 0} \frac{i e^2 \, (k_1 \cdot \chi)^2}{(s-m^2) (u-m^2)}
e^{- |{\cal Z}|^2 + \frac{\rho \cdot \chi}{k_1 \cdot \chi}}  + {\cal O}(a^3)
\eea
where the transverse spacelike vector~$a^\mu$, with $k_1\cdot a=0$, is called the ring radius whose magnitude $|a|=\sqrt{a_\mu a^\mu}$ in the case of a Kerr black hole would correspond to the size of the ring singularity. Next, let us introduce four spin-dependent, helicity-independent and dimensionless variables~\cite{Cangemi:2023ysz}:
\bea 
\label{eq:clComptonVars}
x &=& a \cdot (q_2-q_4) \,, \qquad\;\,\qquad
y = a \cdot (q_2+q_4)\,, 
\nn\\
z &=& |a| \frac{k_1 \cdot (q_2-q_4)}{m} \,, \qquad \quad
w = \frac{(a \cdot \chi)\;[k_1 \cdot (q_2-q_4)]}{k_1\cdot\chi} \,.
\eea
Using that $|{\cal Z}|^2 = 2 m |a|$ and the classical limit of $\rho \cdot \chi$ discussed in Ref.~\cite{Cangemi:2023ysz}, we find that
\beq
\lim_{\hbar \to 0} e^{- |{\cal Z}|^2 + \frac{\rho \cdot \chi}{k_1 \cdot \chi}}
= e^{x-w} .
\eeq
Therefore, using the fact that $k_1 \cdot (q_2 + q_4) \sim \hbar^2$ and hence we can take 
$[ k_1 \cdot (q_2 + q_4) ]^2 \Big|_{\hbar \to 0} \approx 0$, we finally obtain the classical amplitude valid up to $a^3$:
\beq
A_{\textrm{cl}}({\bf 1},2^{+}, {\bf 3}, 4^{-}) =  i e^2 
\frac{ (k_1 \cdot \chi)^2}{ [ k_1 \cdot (q_2 - q_4) ]^2}
e^{x} \left( 1 - w + \frac{w^2}{2} \right) + {\cal O}(a^3) .
\eeq
We expand this way since $w$ can only emerge at most quadratically so that it will not generate the spurious pole discussed in the literature. Therefore, comparing with our amplitude for generic integer $S$, we learn that
\beq
\lim_{\hbar \to 0} e^{- |{\cal Z}|^2} ( \rho \cdot \chi)^{2}
= e^{x} \left( 1 - w + \frac{w^2}{2} \right) .
\eeq
We have to be careful when inserting this into our amplitude in order to avoid double counting the lowest order terms in the expansion in powers of $a$. In other words, the contribution for $n=1$ should be part of the low-orders expansion of the exponential $e^{x-w}$. This can be easily achieved by isolating the $n=1$ term from the rest and taking due care of the expansion of $e^{x-w}$ so that we have at most quadratic orders in $w$ and absence of the aforementioned double counting. In this way, we obtain 
\begingroup
\allowdisplaybreaks
\bea
A_{\textrm{cl}}({\bf 1},2^{+}, {\bf 3}, 4^{-}) &=& 
i e^2 
\frac{ (k_1 \cdot \chi)^2}{ [ k_1 \cdot (q_2 - q_4) ]^2} \Bigl[1 + ( e^{x} - 1) \Bigr]
\left[ 1 + \left( - w + \frac{w^2}{2} \right) \right]
\nn\\
&+& i e^2 
\frac{ (k_1 \cdot \chi)^2}{ [ k_1 \cdot (q_2 - q_4) ]^2}
\lim_{\hbar \to 0}  \sum_{n=2}^{\infty} \frac{1}{(2 (n-1))!} 
\left(  \frac{\sqrt{\pi } \Gamma(n)}{ 4^{n-1} \Gamma\left(n-\frac{1}{2}\right) } \right)
(-1)^{n-1}
\nn\\
&\times& \left\{ \left( \frac{w^2}{2} - w \right) ( e^x - 1)
\left[ \left( \frac{ V^2 - \Bigl( 4 W_{-} - (U + m  [ {\bf 1} {\bf 3} ]) \Bigr)^2 }{2 m^4} \right)^{n-1} 
\right. \right.
\nn\\
&\times& \left. \left.
  \, _2F_1\left(1-n,1-n;1; \frac{\Bigl( V^2 
  - (U + m  [ {\bf 1} {\bf 3} ])^2 \Bigr)}
 {V^2 - \Bigl( 4 W_{-} - (U + m  [ {\bf 1} {\bf 3} ]) \Bigr)^2}\right)
\right. \right.
 \nn\\
&+& \left. \left.
\left( \frac{ V^2 - \Bigl( 4 W_{-} + U + m  \langle {\bf 1} {\bf 3} \rangle \Bigr)^2 }{2 m^4} \right)^{n-1} 
\right. \right.
\nn\\
&\times& \left. \left.
  \, _2F_1\left(1-n,1-n;1; \frac{\Bigl( V^2 
  - (U + m  \langle {\bf 1} {\bf 3} \rangle)^2 \Bigr)}
 {V^2 - \Bigl( 4 W_{-} + U + m  \langle {\bf 1} {\bf 3} \rangle \Bigr)^2}\right) \right]
\right.
\nn\\
&+& \left.
\left( \frac{w^2}{2} - w \right)
\left[ \left( \frac{ V^2 - \Bigl( 4 W_{-} - (U + m  [ {\bf 1} {\bf 3} ]) \Bigr)^2 }{2 m^4} \right)^{n-1} 
\right. \right.
\nn\\
&\times& \left. \left.
  \, _2F_1\left(1-n,1-n;1; \frac{\Bigl( V^2 
  - (U + m  [ {\bf 1} {\bf 3} ])^2 \Bigr)}
 {V^2 - \Bigl( 4 W_{-} - (U + m  [ {\bf 1} {\bf 3} ]) \Bigr)^2}\right)
\right. \right.
 \nn\\
&+& \left. \left.
\left( \frac{ V^2 - \Bigl( 4 W_{-} + U + m  \langle {\bf 1} {\bf 3} \rangle \Bigr)^2 }{2 m^4} \right)^{n-1} 
\right. \right.
\nn\\
&\times& \left. \left.
  \, _2F_1\left(1-n,1-n;1; \frac{\Bigl( V^2 
  - (U + m  \langle {\bf 1} {\bf 3} \rangle)^2 \Bigr)}
 {V^2 - \Bigl( 4 W_{-} + U + m  \langle {\bf 1} {\bf 3} \rangle \Bigr)^2}\right) \right]
\right.
\nn\\
&+& \left.
( e^x - 1)
\left[ \left( \frac{ V^2 - \Bigl( 4 W_{-} - (U + m  [ {\bf 1} {\bf 3} ]) \Bigr)^2 }{2 m^4} \right)^{n-1} 
\right. \right.
\nn\\
&\times& \left. \left.
  \, _2F_1\left(1-n,1-n;1; \frac{\Bigl( V^2 
  - (U + m  [ {\bf 1} {\bf 3} ])^2 \Bigr)}
 {V^2 - \Bigl( 4 W_{-} - (U + m  [ {\bf 1} {\bf 3} ]) \Bigr)^2}\right)
\right. \right.
 \nn\\
&+& \left. \left.
\left( \frac{ V^2 - \Bigl( 4 W_{-} + U + m  \langle {\bf 1} {\bf 3} \rangle \Bigr)^2 }{2 m^4} \right)^{n-1} 
\right. \right.
\nn\\
&\times& \left. \left.
  \, _2F_1\left(1-n,1-n;1; \frac{\Bigl( V^2 
  - (U + m  \langle {\bf 1} {\bf 3} \rangle)^2 \Bigr)}
 {V^2 - \Bigl( 4 W_{-} + U + m  \langle {\bf 1} {\bf 3} \rangle \Bigr)^2}\right) \right]
\right\} .
\eea
\endgroup
The sum over the coherent states will produce an entire function which can be expanded according to the discussion in Ref.~\cite{Cangemi:2023ysz}. Hence the final result for the classical opposite-helicity amplitude reads
\bea
A_{\textrm{cl}}({\bf 1},2^{+}, {\bf 3}, 4^{-}) &=& 
i e^2 \frac{ (k_1 \cdot \chi)^2}{ [ k_1 \cdot (q_2 - q_4) ]^2}
\Bigl[1 + ( e^{x} - 1) \Bigr]
\left[ 1 + \left( - w + \frac{w^2}{2} \right) \right]
\nn\\ 
&+& i e^2 
\frac{ (k_1 \cdot \chi)^2}{ [ k_1 \cdot (q_2 - q_4) ]^2}
\sum_{n=2}^{\infty} \frac{1}{(2 (n-1))!} 
\left(  \frac{\sqrt{\pi } \, \Gamma(n)}{ 4^{n-1} \Gamma\left(n-\frac{1}{2}\right) } \right)
\nn\\
&\times& \left\{ 
( e^{x} - 1) \left( \frac{w^2}{2} - w \right)
\left[ \Bigl( ({x}+ 5{y})^2 - {z}^2 \Bigr)^{n-1}
\, _2F_1\left(1-n,1-n;1; \frac{({x}+{y})^2-{z}^2}{({x}+ 5{y})^2 - {z}^2} \right)
\right. \right.
\nn\\
&+& \left. \left.
\Bigl( ({x} - 5{y})^2 - {z}^2 \Bigr)^{n-1}
\, _2F_1\left(1-n,1-n;1; \frac{({x}-{y})^2-{z}^2}{({x} - 5{y})^2 - {z}^2} \right) \right]
\right.
\nn\\
&+& \left. \left( \frac{w^2}{2} - w \right)
\left[ \Bigl( ({x}+ 5{y})^2 - {z}^2 \Bigr)^{n-1}
\, _2F_1\left(1-n,1-n;1; \frac{({x}+{y})^2-{z}^2}{({x}+ 5{y})^2 - {z}^2} \right)
\right. \right.
\nn\\
&+& \left. \left.
\Bigl( ({x} - 5{y})^2 - {z}^2 \Bigr)^{n-1}
\, _2F_1\left(1-n,1-n;1; \frac{({x}-{y})^2-{z}^2}{({x} - 5{y})^2 - {z}^2} \right) \right]
\right.
\nn\\
&+& \left. ( e^{x} - 1) 
\left[ \Bigl( ({x}+ 5{y})^2 - {z}^2 \Bigr)^{n-1}
\, _2F_1\left(1-n,1-n;1; \frac{({x}+{y})^2-{z}^2}{({x}+ 5{y})^2 - {z}^2} \right)
\right. \right.
\nn\\
&+& \left. \left.
\Bigl( ({x} - 5{y})^2 - {z}^2 \Bigr)^{n-1}
\, _2F_1\left(1-n,1-n;1; \frac{({x}-{y})^2-{z}^2}{({x} - 5{y})^2 - {z}^2} \right) \right]
 \right\}
 + {\cal O}(\hbar) .
\eea
Even though the appearance of the hypergeometric function makes the result sound a little sophisticated, it can be easily shown that the function actually reduces to a polynomial. Hence the final result is completely expressed in terms of entire functions. 

Expanding the result up to ${\cal O}(a^3)$, our results agree with those in Ref.~\cite{Cangemi:2023ysz}. At higher orders, our results seem to differ. Amusingly, setting $y=0$, we find the same operators in the expansion in powers of $a$ (i.e. powers of $x$ and $z$), albeit accompanied by different coefficients (at least up to ${\cal O}(x^6)$ and ${\cal O}(z^6)$). On the other hand, the $n=1$ case satisfies the so-called spin-shift symmetry~\cite{Bautista:2022wjf,Aoude:2022trd,Bern:2022kto}, $a_i^{\mu} \to a_i^{\mu} + \iota_i q^{\mu}/q^2$, where $q$ is the momentum transfer and $\iota_i$ is an arbitrary parameter. As emphasized by Ref.~\cite{Aoude:2022trd}, the emergence of lone factors of $q \cdot a$ will violate the spin-shift symmetry. We conclude that the classical limit of our proposal for the opposite-helicity amplitude should fail to obey this symmetry for $n>1$.

\section{Gravitational Compton amplitudes for all spins}

\subsection{Double copy and first mass level}

Now let us study the gravity Compton amplitude. Using our previous prescription for deriving the associated field-theory amplitudes, we get
\beq
M^{\alpha\gamma,\beta\delta}_{4}({\bf 1}, 2, {\bf 3},4) = 64 i \left( \frac{(-1)^{n}}{n [\Gamma(n)]^4} \right) \kappa^2 m^2  
\frac{1}{\alpha^{\prime} t  ( s \alpha^{\prime} - 4n ) ( u \alpha^{\prime} - 4n )} 
\left[ {\cal A}_0^{\alpha\beta} {\cal A}_0^{\gamma\delta} \right]_{\alpha^{\prime} \to \alpha^{\prime}/4}
\label{43} 
\eeq
where we have expanded $\sin\left( \pi x \right)$ around $x=n$ and also restored the gravity coupling constants, $M_N \to \kappa^{N-2} M_N $~\cite{BCJ1} ($N$ is the number of external particles), where $\kappa^2 = 32 \pi G$. The first mass level $n=1$ corresponds in our prescription to a massive spin-2 particle since here we are compactifying two polarizations of each of the gauge amplitudes. In turn, it is easy to see that our expression agrees with the one derived in Ref.~\cite{Bautista:2022wjf} in quadratic order in $a^{\mu}$, as expected from the KLT formula, see also Ref.~\cite{Bautista:2019tdr}. In fact, up to fifth order in classical spin our result for the opposite helicity amplitude also matches with the one given in Ref.~\cite{Bautista:2022wjf}, as we will show below. For opposite helicities, we can write that
\bea
M_{S}({\bf 1},2^{++}, {\bf 3}, 4^{--}) &=& - i \frac{\kappa^2}{4 \, m^{2S}} 
\frac{\langle 4 | {\bf 1} | 2 \bigr]^{4}}{(s-m^2) (u-m^2) t}
\langle {\bf 3} |^{2S} \exp\left( i \frac{q_4^{\mu} \bar{\epsilon}_{4}^{\nu} {\cal J}_{1 \mu\nu}}
{\bar{p} \cdot \bar{\epsilon}_{4}} \right)  | {\bf 1} \rangle^{2S}
\nn\\
&=& - i \frac{\kappa^2}{4} \frac{1}{(s-m^2) (u-m^2) t}
\langle 4| {\bf 1} |2 \bigl]^{4-2S}
\Bigl( \langle {\bf 3} 4 \rangle \bigl[ {\bf 1} 2 \bigr] + \langle {\bf 1} 4 \rangle \bigl[ {\bf 3} 2 \bigr] \Bigr)^{2S}
\,\,\,\,\,
(S \leq 2)
\eea
which is just the standard result~\cite{Guevara:19a,Johansson:19}. As above, one can also perform the same analysis for the same-helicity amplitude, finding the usual value quoted in the literature (see the general expression below).

\subsection{Double copy and all mass levels}

Let us now present the gravitational Compton amplitude for even positive integer $S$. After imposing the three constraints listed above, the KLT relations allows us to derive our candidate 
opposite helicity gravity Compton amplitude which is valid for all quantum spins $S = 2n>0$:
\bea
\bar{\cal M}_{S}({\bf 1}, 2^{++}, {\bf 3},4^{--}) &=& 64 i \left( \frac{(-1)^{n}}{n [\Gamma(n)]^4} \right)  \kappa^2 m^2
\frac{1}{ \alpha^{\prime} t ( \alpha^{\prime} s - 4 n ) ( \alpha^{\prime} u - 4 n ) } 
\nn\\
&\times& \left[  \alpha^{\prime 4} \mathbb{A}_{S/2}({\bf 1}, 2^{+}, {\bf 3}, 4^{-})
\mathbb{A}_{S/2}({\bf 1}, 2^{+}, {\bf 3}, 4^{-}) \right]_{\alpha^{\prime} \to \alpha^{\prime}/4} .
\label{48}
\eea
One can also similarly construct the all plus gravitational amplitude from the corresponding electromagnetic case and the result matches the one known in the literature~\cite{Johansson:19}
\beq
\bar{\cal M}_{S}({\bf 1},2^{++}, {\bf 3}, 4^{++}) = - i m^{4-2S} \frac{\kappa^2}{4 }  
\frac{\bigl[ 24 \bigr]^4 }{t}
\frac{\langle {\bf 1} {\bf 3} \rangle^{2S}}{[s - m^2] [u - m^2]}, 
\eeq
apart from a positive multiplicative numerical factor (not shown in the expression above). This formula is valid to all $S$. 

Let us see how the 3-point Kerr amplitude can also be identified as a contribution in the full closed string amplitude, after imposing the above constraints. We simply repeat the calculation for the closed superstring exploiting the simple  KLT relation at three points
\beq
M_3({\bf 1}_S, {\bf 2}_S, k) = i \kappa
\left[ \frac{1}{e} A_3({\bf 1}_{S/2}, {\bf 2}_{S/2}, k)  \right]	
\left[ \frac{1}{e} A_3({\bf 1}_{S/2}, {\bf 2}_{S/2}, k) \right],
\eeq
where as above we perform the rescaling $\alpha^{\prime} \to \alpha^{\prime}/4$ which is counterbalanced by the fact that the leading Regge closed string states produce $\alpha^{\prime} = 2 (S-2)/m^2$. We find that
\bea
M_3({\bf 1}, {\bf 2}, k^{++}) &=& (-1)^{S} \frac{i \kappa}{2} (m x_1^+)^2 
\left( \frac{\langle {\bf 1} {\bf 2} \rangle}{m} \right)^{2S}
\Bigl( 1 + {\cal O}(1/m) \Bigr)
\nn\\
M_3({\bf 1}, {\bf 2}, k^{--}) &=& (-1)^{S} \frac{i \kappa}{2} (m x_1^-)^2 
\left( \frac{\bigl[ {\bf 1} {\bf 2} \bigr]}{m} \right)^{2S}
\Bigl( 1 + {\cal O}(1/m) \Bigr)
\eea
which of course is the usual result for the minimally coupled gravitational amplitude.

Similar to the electromagnetic case, our claim here is that it is possible that closed string-theory amplitudes secretly enclose compact-body amplitudes. Furthermore, observe that we impose the constraints on the gravitational amplitude after performing the double copy procedure on the full string amplitude. This could make the double-copy construction compatible with factorization poles that match the known 3-point amplitudes. We believe this happens because our amplitude inherits some aspects of the full original string amplitude; it is a ``stringy" quantum-field-theory amplitude.

\subsection{Classical limit of the opposite-helicity gravitational amplitude}

Here we will present the classical limit of Eq.~(\ref{48}). Again, as discussed above there are many works in the literature concerning the calculation of this amplitude, but for practical purposes we choose to use Ref.~\cite{Cangemi:2023bpe} as our main source of comparison. Many details that will not be exhibited here can be found in such a work. Again we start by rewriting the amplitude in terms of the variables $W_{\pm}, U, V, \rho$ and $\chi$. We find that
\begingroup
\allowdisplaybreaks
\bea
\bar{\cal M}_{S}({\bf 1}, 2^{++}, {\bf 3},4^{--}) &=& i \left( \frac{(-1)^{n}}{ [\Gamma(n)]^4} \right)  \kappa^2 
\frac{1}{  t ( s - m^2 ) ( u - m^2 ) } 
\nn\\
&\times& \left( {\cal N} \frac{n^2 (2 n)^{n-1} (-1)^{n-1}}{2 \, \Gamma (n)^2}
\gamma(s \alpha^{\prime}/4,n-1) \gamma(u \alpha^{\prime}/4,n-1) \right)^2
( \rho \cdot \chi)^{4}
\nn\\
&\times&
\left[ \left( \frac{ V^2 - \Bigl( 4 W_{-} - (U + m  [ {\bf 1} {\bf 3} ]) \Bigr)^2 }{2 m^4} \right)^{n-1} 
\right.
\nn\\
&\times& \left.
  \, _2F_1\left(1-n,1-n;1; \frac{\Bigl( V^2 
  - (U + m  [ {\bf 1} {\bf 3} ])^2 \Bigr)}
 {V^2 - \Bigl( 4 W_{-} - (U + m  [ {\bf 1} {\bf 3} ]) \Bigr)^2}\right)
\right.
 \nn\\
&+& \left. 
\left( \frac{ V^2 - \Bigl( 4 W_{-} + U + m  \langle {\bf 1} {\bf 3} \rangle \Bigr)^2 }{2 m^4} \right)^{n-1} 
\right.
\nn\\
&\times& \left.
  \, _2F_1\left(1-n,1-n;1; \frac{\Bigl( V^2 
  - (U + m  \langle {\bf 1} {\bf 3} \rangle)^2 \Bigr)}
 {V^2 - \Bigl( 4 W_{-} + U + m  \langle {\bf 1} {\bf 3} \rangle \Bigr)^2}\right)
\right]^2
\eea
where we recall that $S = 2n$. The classical limit is obtained with
\bea
M_{\textrm{cl}}({\bf 1}, 2^{++}, {\bf 3},4^{--}) &=& \lim_{\hbar \to 0}
\frac{i \, \kappa^2}{4} \frac{1}{  t ( s - m^2 ) ( u - m^2 ) } e^{- |{\cal Z}|^2} ( \rho \cdot \chi)^{4}
\nn\\
&\times& 
\sum_{n=1}^{\infty} \frac{1}{(2 (n-1))!} 
\left(  \frac{\sqrt{\pi } }{ 4^{n-1} \Gamma(n) \Gamma\left(n-\frac{1}{2}\right) } \right)^2
(-1)^{3n}
\nn\\
&\times&
\left[ \left( \frac{ V^2 - \Bigl( 4 W_{-} - (U + m  [ {\bf 1} {\bf 3} ]) \Bigr)^2 }{2 m^4} \right)^{n-1} 
\right.
\nn\\
&\times& \left.
  \, _2F_1\left(1-n,1-n;1; \frac{\Bigl( V^2 
  - (U + m  [ {\bf 1} {\bf 3} ])^2 \Bigr)}
 {V^2 - \Bigl( 4 W_{-} - (U + m  [ {\bf 1} {\bf 3} ]) \Bigr)^2}\right)
\right.
 \nn\\
&+& \left. 
\left( \frac{ V^2 - \Bigl( 4 W_{-} + U + m  \langle {\bf 1} {\bf 3} \rangle \Bigr)^2 }{2 m^4} \right)^{n-1} 
\right.
\nn\\
&\times& \left.
  \, _2F_1\left(1-n,1-n;1; \frac{\Bigl( V^2 
  - (U + m  \langle {\bf 1} {\bf 3} \rangle)^2 \Bigr)}
 {V^2 - \Bigl( 4 W_{-} + U + m  \langle {\bf 1} {\bf 3} \rangle \Bigr)^2}\right)
\right]^2 .
\eea
\endgroup
To understand the $\hbar \to 0$ limit of the term $e^{- |{\cal Z}|^2} ( \rho \cdot \chi)^{4}$, we again resort to the standard expression for $M_{S}({\bf 1},2^{++}, {\bf 3}, 4^{--})$ with $S \leq 2$ given above. After performing similar steps as the ones carried out in the previous section, we find that 
\bea
M_{S=2, \textrm{cl}}({\bf 1},2^{++}, {\bf 3}, 4^{--}) = - i \frac{\kappa^2}{4} 
\frac{ (k_1 \cdot \chi)^4}{ [ k_1 \cdot (q_2 - q_4) ]^2 (q_2 + q_4)^2}
e^{x} \left( 1 - w + \frac{w^2}{2} - \frac{w^3}{3!} + \frac{w^4}{4!} \right) + {\cal O}(a^5) .
\eea
where again the expansion follows from the fact that we can have at most $w^4$ to prevent spurious poles from showing up. We can use this result in our amplitude and, as before, we prevent possible double-counting by suitably expanding $e^{x-w}$. In this way we obtain our final formula for the classical gravitational amplitude for opposite helicities:
\bea
M_{\textrm{cl}}({\bf 1}, 2^{++}, {\bf 3},4^{--}) &=& - \frac{i \, \kappa^2}{4} 
\frac{ (p_1 \cdot \chi)^4}{ [ p_1 \cdot (q_2 - q_4) ]^2 (q_2 + q_4)^2}
[ 1 + ( e^{x} - 1 ) ] \left[ 1 + \left( - w + \frac{w^2}{2} - \frac{w^3}{3!} + \frac{w^4}{4!} \right) \right]
\nn\\
&+& \frac{i \, \kappa^2}{4} 
\frac{ (p_1 \cdot \chi)^4}{ [ p_1 \cdot (q_2 - q_4) ]^2 (q_2 + q_4)^2}
\sum_{n=2}^{\infty} \frac{1}{(2 (n-1))!} 
\left(  \frac{\sqrt{\pi } }{ 4^{n-1} \Gamma(n) \Gamma\left(n-\frac{1}{2}\right) } \right)^2
(-1)^{n}
\nn\\
&\times& \left\{ 
( e^{x} - 1)  \left( - w + \frac{w^2}{2} - \frac{w^3}{3!} + \frac{w^4}{4!} \right)
\right.
\nn\\
&\times& \left. \left[ \Bigl( ({x}+ 5{y})^2 - {z}^2 \Bigr)^{n-1}
\, _2F_1\left(1-n,1-n;1; \frac{({x}+{y})^2-{z}^2}{({x}+ 5{y})^2 - {z}^2} \right)
\right. \right.
\nn\\
&+& \left. \left.
\Bigl( ({x} - 5{y})^2 - {z}^2 \Bigr)^{n-1}
\, _2F_1\left(1-n,1-n;1; \frac{({x}-{y})^2-{z}^2}{({x} - 5{y})^2 - {z}^2} \right) \right]^2
\right.
\nn\\
&+& \left.  \left( - w + \frac{w^2}{2} - \frac{w^3}{3!} + \frac{w^4}{4!} \right)
\left[ \Bigl( ({x}+ 5{y})^2 - {z}^2 \Bigr)^{n-1}
\, _2F_1\left(1-n,1-n;1; \frac{({x}+{y})^2-{z}^2}{({x}+ 5{y})^2 - {z}^2} \right)
\right. \right.
\nn\\
&+& \left. \left.
\Bigl( ({x} - 5{y})^2 - {z}^2 \Bigr)^{n-1}
\, _2F_1\left(1-n,1-n;1; \frac{({x}-{y})^2-{z}^2}{({x} - 5{y})^2 - {z}^2} \right) \right]^2
\right.
\nn\\
&+& \left. ( e^{x} - 1) 
\left[ \Bigl( ({x}+ 5{y})^2 - {z}^2 \Bigr)^{n-1}
\, _2F_1\left(1-n,1-n;1; \frac{({x}+{y})^2-{z}^2}{({x}+ 5{y})^2 - {z}^2} \right)
\right. \right.
\nn\\
&+& \left. \left.
\Bigl( ({x} - 5{y})^2 - {z}^2 \Bigr)^{n-1}
\, _2F_1\left(1-n,1-n;1; \frac{({x}-{y})^2-{z}^2}{({x} - 5{y})^2 - {z}^2} \right) \right]^2
 \right\}
 + {\cal O}(\hbar) .
\eea
Expanding the result up to ${\cal O}(a^5)$, our results agree with those in Ref.~\cite{Cangemi:2023bpe}. At higher orders, our results differ. Moreover, the classical limit of the opposite-helicity amplitude does not satisfy the spin-shift symmetry for quantum spins greater than 2; in any case, the massive-spin 2 opposite-helicity amplitude ($S=2$) obtained from the above double copy formula preserves this symmetry. 

\section{Conclusions and perspectives}

In this work we have proposed candidate Compton amplitudes which are valid to all integer spins.  An amusing feature is that the total amplitude~(\ref{26}) contains both Comptons for same and opposite helicities of photons (and gravitons, in the gravitational case, Eq.~(\ref{43})). This observation is somewhat reminiscent of the Compton amplitude written in terms of massive spinors in the high-energy limit~\cite{Arkani-Hamed:2017jhn}; for instance, for the massive spin-one amplitude, the high-energy limit is given by a sum of its ``helicity constituents''. As helicity amplitudes the sum of components is generally not allowed, yet this is precisely how one can sort out the different pieces of the amplitude that unifies all the different helicity amplitudes merged into a single object. Remarkably, we observe a similar phenomenon in the present case. 

There are other prominent features of such amplitudes which potentially justify the given interpretation. We list these here: 

\begin{enumerate}

\item Amplitudes are gauge invariant; it is possible to rewrite them in terms of gauge invariant operators;

\item Amplitudes do not present any spurious poles and hence they are valid to all spins;

\item We have found no evidence of the general validity of the spin-shift symmetry;

\item By construction, it has the correct factorization in the physical channels;

\item The gravitational quantum amplitudes obey double copy. 

\end{enumerate}

As we have shown in the case of the first excited state $n=1$, the compactification down to the spin-$1$ case allowed us to recover known results in the literature and supplied the motivation to search for a valid amplitude for higher spins based on a string-theory calculation. We argued that this works also for the case of the 3-point amplitude. Nevertheless, it should be noted that this is certainly an ambiguous procedure (at best) that calls for a better comprehension. One possible conclusion is that subleading Regge trajectories could be the key to understanding why this kind of exotic compactification is needed here.

While comparisons to other works could certainly be enlightening, we chose to compare our classical Compton-amplitude results to those of Refs.~\cite{Cangemi:2023ysz,Cangemi:2023bpe}, as we have emphasized above. The reason for this is that these works allow a direct parallel with our results, facilitating not only the extraction of the classical limit of the amplitudes presented here, but also making clear the physics behind the outcomes. Indeed, one can even speculate whether there is a UV completion within the framework of string theory for the massive higher-spin effective theory investigated in such references. On the other hand, our soft expansion requires some refinement, perhaps following standard procedures in the literature~\cite{DiVecchia:2015srk,Sen:2017xjn,DiVecchia:2019kle}. Going further, the worldsheet interpretation for Kerr provided by Ref.~\cite{worldsheetpaper} might fit well with our description. In turn, as mentioned above, to obtain the corresponding gravitational amplitudes we not only correlate the helicites in the gauge amplitudes, but also the spins of the massive particles. So it is difficult to discuss the independence of the amplitude on the split $S \to S_1+S_2$ within our formalism. This also motivates the study of the R sector of the superstring. We intend to explore such research directions in the future.

\vspace{-0.2cm}

\section*{Acknowledgements}

We thank Paolo di Vecchia and Francesco Alessio for suggestions and comments on the draft, and for sharing with us their notes with similar calculations and results. We also thank Yilber Fabian Bautista, Radu Roiban, Zvi Bern, N. Emil J. Bjerrum-Bohr, Renann Jusinskas, Rafael Aoude, Riccardo Gonzo, Matteo Sergola, Henrik Johansson and Fei Teng for their invaluable remarks and suggestions. GM acknowledges partial support from FAPESP under grant 2023/06508-8, CNPq under grant 317548/2021-2 and FAPERJ under grant E-26/201.142/2022.

\section*{Appendix A. Details of the formula for the Compton scattering amplitude}

Here we present the explicit for the quantity ${\cal A}_0^{\alpha\beta}$. The amplitude reads
\beq
A^{\alpha\beta}_4({\bf 1}, 2, {\bf 3},4) \equiv A^{\alpha\beta}_t = 
\frac{{\cal V}_t}{\alpha^{\prime} t \, \prod_{k=1}^{2n-1} (\alpha^{\prime} t - k)}
{\cal A}_0^{\alpha\beta}
\eeq
where we have written
\beq
{\cal A}_0^{\alpha\beta} \equiv {\cal A}^{\alpha\beta (1)}_4
+ {\cal A}^{\alpha\beta (2)}_4
+ {\cal A}^{\alpha\beta (3)}_4
+ {\cal A}^{\alpha\beta (4)}_4 
\eeq
with
\begingroup
\allowdisplaybreaks
\bea
{\cal A}^{\alpha\beta (1)}_4({\bf 1}, 2, {\bf 3},4) &=& 2 i \alpha^{\prime} e^2(n!)^2
 \sum_{j=0}^{n} \frac{(2 \alpha^{\prime})^{n-j}}{j! [(n-j)!]^2} 
\eta^{a(j) c(j)}
\nn\\
&\times&
\sum_{k=0}^{n-j}
\begin{pmatrix}
n-j \\
k
\end{pmatrix}
(-1)^{k} q_2^{a(n-j-k)} q_4^{a(k)}
\sum_{l=0}^{n-j}
\begin{pmatrix}
n-j \\
l
\end{pmatrix}
(-1)^{n-j-l} q_2^{c(n-j-l)} q_4^{c(l)}
\nn\\
&\times& \biggl\{
\frac{1}{(\alpha^{\prime} t+1)}
\gamma(s \alpha^{\prime},k-l+n) \gamma(u \alpha^{\prime},-k+l+n)
\eta^{a_S c_S}
\left( q_2^{\beta} q_4^{\alpha} - (q_2 \cdot q_4) \eta^{\alpha\beta}  \right)
\nn\\
&+& \Bigl[ (-u \alpha^{\prime}-k+l+n)
 \left( q_2^{a_S}
\left(  q_4^{c_S} \eta^{\alpha \beta} - q_4^{\alpha} \eta^{c_S \beta}  \right)
- \eta^{a_S \alpha} \left(  q_2^{\beta} q_4^{c_S} - (q_2 \cdot q_4) \eta^{c_S \beta} \right) \right)
\nn\\
&-& (-s \alpha^{\prime}+k-l+n)
\left( q_4^{a_S}
\left(  q_2^{\beta} \eta^{\alpha c_S}  - q_2^{c_S} \eta^{\alpha\beta}  \right)
- \eta^{a_S \beta}
\left(  (q_2 \cdot q_4) \eta^{\alpha c_S}  - q_2^{c_S} q_4^{\alpha}  \right) \right) \Bigr]
\nn\\
&\times& \gamma(s \alpha^{\prime},k-l+n-1) \gamma(u \alpha^{\prime},-k+l+n-1)
\biggr\}
\eea
\bea
{\cal A}^{\alpha\beta (2)}_4({\bf 1}, 2, {\bf 3},4) &=& 2 i \alpha^{\prime} e^2(n!)^2
\sum_{j=0}^{n} 
\frac{(2 \alpha^{\prime})^{n-j}}{j! [(n-j)!]^2} 
\eta^{a(j) c(j)}
\left( q_2^{a_S} \eta^{\alpha c_S} - q_2^{c_S} \eta^{a_S \alpha} \right)
\nn\\
&\times&
\biggl\{
- \sum_{k=0}^{n-j}
\begin{pmatrix}
n-j \\
k
\end{pmatrix}
(-1)^{k} q_2^{a(n-j-k)} q_4^{a(k)}
\sum_{l=0}^{n-j}
\begin{pmatrix}
n-j \\
l
\end{pmatrix}
(-1)^{n-j-l} q_2^{c(n-j-l)} q_4^{c(l)}
\nn\\
&\times& \left( (-s \alpha^{\prime}+k-l+n) q_2^{\beta} +  \alpha^{\prime} t \,  k_3^{\beta}  \right) 
\gamma(s \alpha^{\prime},k-l+n-1)\gamma(u \alpha^{\prime},-k+l+n-1)
\nn\\
&+&  \frac{(n-j) \alpha^{\prime} t}{(2 \alpha^{\prime})} 
\sum_{k=0}^{n-j-1}
\begin{pmatrix}
n-j-1 \\
k
\end{pmatrix}
(-1)^{k} q_2^{a(n-j-1-k)} q_4^{a(k)}
\nn\\
&\times&
\sum_{l=0}^{n-j-1}
\begin{pmatrix}
n-j-1 \\
l
\end{pmatrix}
(-1)^{n-j-1-l} q_2^{c(n-j-1-l)} q_4^{c(l)}
\nn\\
&\times&
 \left[  \eta^{\beta a_n} 
\Bigl(  - \gamma(s \alpha^{\prime},k-l+n) \gamma(u \alpha^{\prime},-k+l+n-2) q_2^{c_{n}} 
\right.
\nn\\
&+& \left. \gamma(s \alpha^{\prime},k-l+n-1) \gamma(u \alpha^{\prime},-k+l+n-1) q_4^{c_{n}}  \Bigr)
\right.
\nn\\
&+& \left.  \eta^{\beta c_n} 
\Bigl( \gamma(s \alpha^{\prime},k-l+n-2) \gamma(u \alpha^{\prime},-k+l+n) q_2^{a_{n}}
\right.
\nn\\
&-& \left. \gamma(s \alpha^{\prime},k-l+n-1) \gamma(u \alpha^{\prime},-k+l+n-1) q_4^{a_{n}}  \Bigr) \right]
\biggr\}
\eea
\bea
{\cal A}^{\alpha\beta (3)}_4({\bf 1}, 2, {\bf 3},4) &=& 2 i \alpha^{\prime} e^2(n!)^2
\sum_{j=0}^{n} 
\frac{(2 \alpha^{\prime})^{n-j}}{j! [(n-j)!]^2} 
\eta^{a(j) c(j)}
\left( q_4^{c_S} \eta^{a_S \beta} - q_4^{a_S} \eta^{c_S \beta}  \right)
\nn\\
&\times&
\left\{
\sum_{k=0}^{n-j}
\begin{pmatrix}
n-j \\
k
\end{pmatrix}
(-1)^{k} q_2^{a(n-j-k)} q_4^{a(k)}
\sum_{l=0}^{n-j}
\begin{pmatrix}
n-j \\
l
\end{pmatrix}
(-1)^{n-j-l} q_2^{c(n-j-l)} q_4^{c(l)}
\right.
\nn\\
&\times& \left.
\Bigl( (-u \alpha^{\prime}-k+l+n) q_4^{\alpha} + \alpha^{\prime} t \, k_3^{\alpha}  \Bigr)
\gamma(s \alpha^{\prime},k-l+n-1) \gamma(u \alpha^{\prime},-k+l+n-1)
\right.
\nn\\
&+& \left. \frac{(n-j) \alpha^{\prime} t }{(2 \alpha^{\prime})} 
\sum_{k=0}^{n-j-1}
\begin{pmatrix}
n-j-1 \\
k
\end{pmatrix}
(-1)^{k} q_2^{a(n-j-1-k)} q_4^{a(k)}
\right.
\nn\\
&\times& \left. \sum_{l=0}^{n-j-1}
\begin{pmatrix}
n-j-1 \\
l
\end{pmatrix}
(-1)^{n-j-1-l} q_2^{c(n-j-1-l)} q_4^{c(l)}
\right.
\nn\\
&\times& \left. \left[  \eta^{\alpha a_n} 
\Bigl( - \gamma(s \alpha^{\prime},k-l+n-1) \gamma(u \alpha^{\prime},-k+l+n-1) q_2^{c_{n}} 
\right. \right.
\nn\\
&+& \left. \left. \gamma(s \alpha^{\prime},k-l+n-2) \gamma(u \alpha^{\prime},-k+l+n) q_4^{c_{n}}  \Bigr)
\right. \right.
\nn\\
&+& \left. \left.  \eta^{\alpha c_n} 
\Bigl( \gamma(s \alpha^{\prime},k-l+n-1) \gamma(u \alpha^{\prime},-k+l+n-1)q_2^{a_{n}}  
\right. \right.
\nn\\
&-& \left. \left. \gamma(s \alpha^{\prime},k-l+n) \gamma(u \alpha^{\prime},-k+l+n-2) q_4^{a_{n}}  \Bigr) \right]
\right\}
\eea
and
\bea
{\cal A}^{\alpha\beta (4)}_4({\bf 1}, 2, {\bf 3},4) &=& 2 i \alpha^{\prime} e^2(n!)^2
\sum_{j=0}^{n} 
\frac{(2 \alpha^{\prime})^{n-j}}{j! [(n-j)!]^2}
\eta^{a(j) c(j)}
\nn\\
&\times& \left\{
- \sum_{k=0}^{n-j}
\begin{pmatrix}
n-j \\
k
\end{pmatrix}
(-1)^{k} q_2^{a(n-j-k)} q_4^{a(k)}
\sum_{l=0}^{n-j}
\begin{pmatrix}
n-j \\
l
\end{pmatrix}
(-1)^{n-j-l} q_2^{c(n-j-l)} q_4^{c(l)}
\right.
\nn\\
&\times& \left. \biggl[ \frac{1}{(\alpha^{\prime} t +1)} 
\gamma(s \alpha^{\prime},k-l+n) \gamma(u \alpha^{\prime},-k+l+n)
q_4^{\alpha} q_2^{\beta}
\right.
\nn\\
&+&  \left.
\Bigl( k_3^{\alpha} \bigl( (-s \alpha^{\prime}+k-l+n)   q_2^{\beta}
+  \alpha^{\prime} t \,  k_3^{\beta} \bigr)
+  (-u \alpha^{\prime}-k+l+n) q_4^{\alpha}  k_3^{\beta} \Bigr) 
\right.
\nn\\
&\times& \left. \gamma(s \alpha^{\prime},k-l+n-1)  \gamma(u \alpha^{\prime},-k+l+n-1) \biggr]
\right.
\nn\\
&-& \left. 
\frac{(n-j) }{(2 \alpha^{\prime})} 
\sum_{k=0}^{n-j-1}
\begin{pmatrix}
n-j-1 \\
k
\end{pmatrix}
(-1)^{k} q_2^{a(n-j-1-k)} q_4^{a(k)}
\right.
\nn\\
&\times& \left. \sum_{l=0}^{n-j-1}
\begin{pmatrix}
n-j-1 \\
l
\end{pmatrix}
(-1)^{n-j-1-l} q_2^{c(n-j-1-l)} q_4^{c(l)}
\right.
\nn\\
&\times& \left.
\biggl[ q_2^{\beta} 
\biggl( \eta^{\alpha a_n} 
\left(  - \gamma(s \alpha^{\prime},k-l+n) \gamma(u \alpha^{\prime},-k+l+n-1) q_2^{c_{n}} 
\right. \right.
\nn\\
&+& \left. \left. \gamma(s \alpha^{\prime},k-l+n-1) \gamma(u \alpha^{\prime},-k+l+n) q_4^{c_{n}} \right)
 \right.
\nn\\
&+&  \left.  \eta^{\alpha c_n} 
\left( \gamma(s \alpha^{\prime},k-l+n) \gamma(u \alpha^{\prime},-k+l+n-1) q_2^{a_{n}} 
\right. \right.
\nn\\
&-& \left. \left. \gamma(s \alpha^{\prime},k-l+n+1) \gamma(u \alpha^{\prime},-k+l+n-2) q_4^{a_{n}} \right) \biggr)
 \right.
\nn\\
&+& \left.
\alpha^{\prime} t \, k_3^{\beta} 
\biggl( \eta^{\alpha a_n} 
\left(  - \gamma(s \alpha^{\prime},k-l+n-1) \gamma(u \alpha^{\prime},-k+l+n-1) q_2^{c_{n}} 
\right. \right.
\nn\\
&+& \left. \left. \gamma(s \alpha^{\prime},k-l+n-2) \gamma(u \alpha^{\prime},-k+l+n) q_4^{c_{n}} \right)
 \right.
\nn\\
&+&  \left.  \eta^{\alpha c_n} 
\left( \gamma(s \alpha^{\prime},k-l+n-1) \gamma(u \alpha^{\prime},-k+l+n-1) q_2^{a_{n}} 
\right. \right.
\nn\\
&-& \left. \left. \gamma(s \alpha^{\prime},k-l+n) \gamma(u \alpha^{\prime},-k+l+n-2) q_4^{a_{n}}  \right) \biggr) \biggr]
\right.
\nn\\
&+& \left. 
\frac{(n-j) }{(2 \alpha^{\prime})} 
\sum_{k=0}^{n-j-1}
\begin{pmatrix}
n-j-1 \\
k
\end{pmatrix}
(-1)^{k} q_2^{a(n-j-1-k)} q_4^{a(k)}
\right.
\nn\\
&\times& \left. \sum_{l=0}^{n-j-1}
\begin{pmatrix}
n-j-1 \\
l
\end{pmatrix}
(-1)^{n-j-1-l} q_2^{c(n-j-1-l)} q_4^{c(l)}
\right.
\nn\\
&\times& \left.
\biggl[  \alpha^{\prime} t \, k_3^{\alpha} 
\Bigl(  \eta^{\beta a_n} 
\left( - \gamma(s \alpha^{\prime},k-l+n) \gamma(u \alpha^{\prime},-k+l+n-2) q_2^{c_{n}} 
\right. \right.
\nn\\
&+& \left. \left. \gamma(s \alpha^{\prime},k-l+n-1) \gamma(u \alpha^{\prime},-k+l+n-1) q_4^{c_{n}} \right)
 \right.
\nn\\
&+&  \left. \eta^{\beta c_n} 
\left( \gamma(s \alpha^{\prime},k-l+n-2) \gamma(u \alpha^{\prime},-k+l+n) q_2^{a_{n}} 
\right. \right.
\nn\\
&-& \left. \left. \gamma(s \alpha^{\prime},k-l+n-1) \gamma(u \alpha^{\prime},-k+l+n-1) q_4^{a_{n}}  \right) \Bigr)
\right.
\nn\\
&+&  \left.
q_4^{\alpha} 
\Bigl(  \eta^{\beta a_n} 
\left( - \gamma(s \alpha^{\prime},k-l+n) \gamma(u \alpha^{\prime},-k+l+n-1) q_2^{c_{n}} 
\right. \right.
\nn\\
&+& \left. \left. \gamma(s \alpha^{\prime},k-l+n-1) \gamma(u \alpha^{\prime},-k+l+n) q_4^{c_{n}} \right)
 \right.
\nn\\
&+&  \left. \eta^{\beta c_n} 
\left(\gamma(s \alpha^{\prime},k-l+n-2) \gamma(u \alpha^{\prime},-k+l+n+1) q_2^{a_{n}} 
\right. \right.
\nn\\
&-& \left. \left. \gamma(s \alpha^{\prime},k-l+n-1) \gamma(u \alpha^{\prime},-k+l+n) q_4^{a_{n}}  \right) \Bigr) \biggr]
\right.
\nn\\
&+& \left. 
\frac{(n-j)(n-j-1)\alpha^{\prime} t  }{(2 \alpha^{\prime})^2}
\sum_{k=0}^{n-j-2}
\begin{pmatrix}
n-j-2 \\
k
\end{pmatrix}
(-1)^{k} q_2^{a(n-j-2-k)} q_4^{a(k)}
\right.
\nn\\
&\times& \left. \sum_{l=0}^{n-j-2}
\begin{pmatrix}
n-j-2 \\
l
\end{pmatrix}
(-1)^{n-j-2-l} q_2^{c(n-j-2-l)} q_4^{c(l)}\right.
\nn\\
&\times& \left.
\biggl[ 
\eta^{\alpha a_n} \eta^{\beta a_{n-1}}
\Bigl( \gamma(s \alpha^{\prime},k-l+n) \gamma(u \alpha^{\prime},-k+l+n-2)
q_2^{c_{n-1}}  q_2^{c_{n}}
 \right.
\nn\\
&-& \left.  \gamma(s \alpha^{\prime},k-l+n-1) \gamma(u \alpha^{\prime},-k+l+n-1)
 q_2^{c_{n-1}}  q_4^{c_{n}}
\right.
\nn\\
&-& \left. \gamma(s \alpha^{\prime},k-l+n-1) \gamma(u \alpha^{\prime},-k+l+n-1)
q_4^{c_{n-1}}  q_2^{c_{n}}
 \right.
\nn\\
&+&  \left. \gamma(s \alpha^{\prime},k-l+n-2) \gamma(u \alpha^{\prime},-k+l+n)
 q_4^{c_{n-1}}  q_4^{c_{n}}
 \Bigr)
 \right.
\nn\\
&+&  \left. \eta^{\alpha c_n} \eta^{\beta c_{n-1}}
\Bigl( \gamma(s \alpha^{\prime},k-l+n-2) \gamma(u \alpha^{\prime},-k+l+n)
q_2^{a_{n-1}}  q_2^{a_{n}} 
 \right.
 \nn\\
&-&   \left. \gamma(s \alpha^{\prime},k-l+n-1) \gamma(u \alpha^{\prime},-k+l+n-1)
  q_2^{a_{n-1}} q_4^{a_{n}} 
\right.
\nn\\
&-& \left. \gamma(s \alpha^{\prime},k-l+n-1) \gamma(u \alpha^{\prime},-k+l+n-1)
 q_4^{a_{n-1}}  q_2^{a_{n}} 
  \right.
\nn\\
&+&  \left. \gamma(s \alpha^{\prime},k-l+n) \gamma(u \alpha^{\prime},-k+l+n-2)
q_4^{a_{n-1}}  q_4^{a_{n}} 
\Bigr)
\biggr]
\right.
\nn\\
&+& \left. 
\frac{(n-j)^2 \alpha^{\prime} t }{(2 \alpha^{\prime})^2}
\sum_{k=0}^{n-j-1}
\begin{pmatrix}
n-j-1 \\
k
\end{pmatrix}
(-1)^{k} q_2^{a(n-j-1-k)} q_4^{a(k)}
\right.
\nn\\
&\times& \left. \sum_{l=0}^{n-j-1}
\begin{pmatrix}
n-j-1 \\
l
\end{pmatrix}
(-1)^{n-j-1-l} q_2^{c(n-j-1-l)} q_4^{c(l)}
\right.
\nn\\
&\times& \left.
\Bigl[ \gamma(s \alpha^{\prime},k-l+n-2) \gamma(u \alpha^{\prime},-k+l+n)
 \eta^{\alpha a_n} \eta^{\beta c_{n}}
\right.
\nn\\
&+& \left. \gamma(s \alpha^{\prime},k-l+n) \gamma(u \alpha^{\prime},-k+l+n-2)
 \eta^{\alpha c_n} \eta^{\beta a_{n}} \Bigr]
\right.
\nn\\
&+& \left. 
\frac{1}{(2 \alpha^{\prime})}
\eta^{\alpha\beta}
\sum_{k=0}^{n-j}
\begin{pmatrix}
n-j \\
k
\end{pmatrix}
(-1)^{k} q_2^{a(n-j-k)} q_4^{a(k)}
\sum_{l=0}^{n-j}
\begin{pmatrix}
n-j \\
l
\end{pmatrix}
(-1)^{n-j-l} q_2^{c(n-j-l)} q_4^{c(l)}
\right.
\nn\\
&\times& \left.
(-s \alpha^{\prime}+k-l+n) (-u \alpha^{\prime}-k+l+n)
\right.
\nn\\
&\times& \left. \frac{1}{(\alpha^{\prime} t +1) } 
\gamma(s \alpha^{\prime},k-l+n-1) \gamma(u \alpha^{\prime},-k+l+n-1)
\right\}
\eta^{a_S c_S} 
\eea
\endgroup
where $\gamma(x,N) \equiv \prod_{m=1}^{N} (-x + m) = (1-x)_N$, $q^{a(m)} \equiv q^{a_1} q^{a_2} \cdots q^{a_m}$, etc., and also we defined $\eta^{a(j) c(j)} \equiv \eta^{a_1 c_1} \eta^{a_2 c_2} \cdots \eta^{a_j c_j}$.  It may appear that there is a tachyonic pole, but if we group all such contributions together:
\bea
&& \frac{1}{(\alpha^{\prime} t+1)}
\left[ \eta^{a_S c_S} \left( q_2^{\beta} q_4^{\alpha} - (q_2 \cdot q_4) \eta^{\alpha\beta}  \right)
- \eta^{a_S c_S} \left( q_2^{\beta} q_4^{\alpha} - \frac{1}{(2 \alpha^{\prime})}
\eta^{\alpha\beta} \right)  \right]
\nn\\
&& = \frac{1}{(\alpha^{\prime} t+1)}
\left[ \eta^{a_S c_S} \left( q_2^{\beta} q_4^{\alpha} 
+ \frac{1}{2 \alpha^{\prime}} \alpha^{\prime} t \eta^{\alpha\beta}  \right)
- \eta^{a_S c_S} \left( q_2^{\beta} q_4^{\alpha} - \frac{1}{(2 \alpha^{\prime})}
\eta^{\alpha\beta} \right)  \right]
\nn\\
&& = \frac{1}{(2 \alpha^{\prime}) (\alpha^{\prime} t+1)}
( \alpha^{\prime} t  +  1 ) \eta^{\alpha\beta}  \eta^{a_S c_S}
= \frac{1}{(2 \alpha^{\prime})} \eta^{\alpha\beta}  \eta^{a_S c_S}
\eea
and so there are no tachyons in the spectrum, as expected. Moreover, the coupling constant $e$ arises by demanding that the amplitude describes the scattering between massive charged particles and photons -- this requirement produces $e^{-\lambda} = \frac{1}{e^2}$.

The other color-ordered amplitudes are given by
\beq
A^{\alpha\beta}_{s} = 
\frac{{\cal V}_s}{\alpha^{\prime} t \, \prod_{k=1}^{2n-1} (\alpha^{\prime} s - k)}
{\cal A}_0^{\alpha\beta}
\eeq
and
\beq
A^{\alpha\beta}_{u} = 
\frac{{\cal V}_u}{\alpha^{\prime} t \, \prod_{k=1}^{2n-1} (\alpha^{\prime} u - k)}
{\cal A}_0^{\alpha\beta}
\eeq
where
\bea
A_t &\equiv& A_4({\bf 1}, 2, {\bf 3}, 4)
\nn\\
A_s &\equiv& A_4(2, {\bf 3}, {\bf 1}, 4)
\nn\\
A_u &\equiv& A_4({\bf 3}, {\bf 1}, 2, 4) .
\eea
To obtain such expressions we made use of the definitions of ${\cal V}_t, \gamma(x,N)$ and repeated application of the recurrence formula
$$
\Gamma(z) = \frac{\Gamma(z+n+1)}{z (z+1) \cdots (z+n)} .
$$
Therefore, the full non-Abelian amplitude can be written as~\cite{Schlotterer:2010kk}
\bea
A_4^{\textrm{full}}({\bf 1}, 2, {\bf 3},4) 
&=& \textrm{tr}\bigl[ T^{a_1} T^{a_2} T^{a_3} T^{a_4} + (-1)^{2n} T^{a_4} T^{a_3} T^{a_2} T^{a_1} \bigr]
A_t
\nn\\
&+& \textrm{tr}\bigl[ T^{a_2} T^{a_3} T^{a_1} T^{a_4} + (-1)^{2n} T^{a_4} T^{a_1} T^{a_3} T^{a_2} \bigr]
A_s
\nn\\
&+& \textrm{tr}\bigl[ T^{a_3} T^{a_1} T^{a_2} T^{a_4} + (-1)^{2n} T^{a_4} T^{a_2} T^{a_1} T^{a_3} \bigr]
A_u 
\eea
where $T^a$ are Chan-Paton matrices, the factor $(-1)^{2n}$ arises due to worldsheet parity. One can use worldsheet monodromy relations,
\bea
\sin(\alpha^{\prime} \pi t) A_s &=& \sin(\alpha^{\prime} \pi s) A_t
\nn\\
\sin(\alpha^{\prime} \pi u) A_t &=& \sin(\alpha^{\prime} \pi t) A_u
\nn\\
\sin(\alpha^{\prime} \pi s) A_u &=& \sin(\alpha^{\prime} \pi u) A_s,
\eea
in order to obtain relations between the three color-ordered partial amplitudes given above. Finally, the momentum kernel ${\cal S}_{\alpha^{\prime}}$ appearing in the KLT formula given in the main text is given by~\cite{Bjerrum-Bohr:2010pnr}
\beq
{\cal S}_{\alpha^{\prime}}[i_1, \ldots, i_k | j_1, \ldots j_k]_{p}
\equiv \left( \frac{\pi \alpha^{\prime}}{4} \right)^{-k}
\prod_{t=1}^{k} \sin\left[ \frac{\pi \alpha^{\prime}}{2} \left( p \cdot k_{i_t} 
+ \sum_{q>t}^{k} \theta(i_t,i_q) k_{i_t} \cdot k_{i_q} \right) \right]
\eeq
where $\theta(i_t,i_q)$ equals 1 if the ordering of the legs $i_t$ and $i_q$ is opposite in the sets 
$\{i_1, \ldots, i_k \}$ and $\{ j_1, \ldots, j_k \}$, and zero if the ordering is the same.

\bibliographystyle{utphys.bst}
\bibliography{references_string.bib}

\providecommand{\href}[2]{#2}\begingroup\raggedright\begin{thebibliography}{100}

\bibitem{Bern:2022wqg}
Z.~Bern, J.~J. Carrasco, M.~Chiodaroli, H.~Johansson, and R.~Roiban, ``{The
  SAGEX review on scattering amplitudes Chapter 2: An invitation to
  color-kinematics duality and the double copy},''
  \href{http://dx.doi.org/10.1088/1751-8121/ac93cf}{{\em J. Phys. A} {\bfseries
  55} no.~44, (2022) 443003}, \href{http://arxiv.org/abs/2203.13013}{{\ttfamily
  arXiv:2203.13013 [hep-th]}}.

\bibitem{KLT}
H.~Kawai, D.~C. Lewellen, and S.~H.~H. Tye, ``{A Relation Between Tree
  Amplitudes of Closed and Open Strings},''
  \href{http://dx.doi.org/10.1016/0550-3213(86)90362-7}{{\em Nucl. Phys. B}
  {\bfseries 269} (1986) 1--23}.

\bibitem{BCJ1}
Z.~Bern, J.~J.~M. Carrasco, and H.~Johansson, ``{New Relations for Gauge-Theory
  Amplitudes},'' \href{http://dx.doi.org/10.1103/PhysRevD.78.085011}{{\em Phys.
  Rev. D} {\bfseries 78} (2008) 085011},
  \href{http://arxiv.org/abs/0805.3993}{{\ttfamily arXiv:0805.3993 [hep-ph]}}.

\bibitem{BCJ2}
Z.~Bern, J.~J.~M. Carrasco, and H.~Johansson, ``{Perturbative Quantum Gravity
  as a Double Copy of Gauge Theory},''
  \href{http://dx.doi.org/10.1103/PhysRevLett.105.061602}{{\em Phys. Rev.
  Lett.} {\bfseries 105} (2010) 061602},
  \href{http://arxiv.org/abs/1004.0476}{{\ttfamily arXiv:1004.0476 [hep-th]}}.

\bibitem{Bern:2019prr}
Z.~Bern, J.~J. Carrasco, M.~Chiodaroli, H.~Johansson, and R.~Roiban, ``{The
  Duality Between Color and Kinematics and its Applications},''
  \href{http://arxiv.org/abs/1909.01358}{{\ttfamily arXiv:1909.01358
  [hep-th]}}.

\bibitem{Brandhuber:2021kpo}
A.~Brandhuber, G.~Chen, G.~Travaglini, and C.~Wen, ``{A new gauge-invariant
  double copy for heavy-mass effective theory},''
  \href{http://dx.doi.org/10.1007/JHEP07(2021)047}{{\em JHEP} {\bfseries 07}
  (2021) 047}, \href{http://arxiv.org/abs/2104.11206}{{\ttfamily
  arXiv:2104.11206 [hep-th]}}.

\bibitem{89}
R.~Monteiro, D.~O'Connell, and C.~D. White, ``{Black holes and the double
  copy},'' \href{http://dx.doi.org/10.1007/JHEP12(2014)056}{{\em JHEP}
  {\bfseries 12} (2014) 056}, \href{http://arxiv.org/abs/1410.0239}{{\ttfamily
  arXiv:1410.0239 [hep-th]}}.

\bibitem{90}
A.~Luna, R.~Monteiro, I.~Nicholson, D.~O'Connell, and C.~D. White, ``{The
  double copy: Bremsstrahlung and accelerating black holes},''
  \href{http://dx.doi.org/10.1007/JHEP06(2016)023}{{\em JHEP} {\bfseries 06}
  (2016) 023}, \href{http://arxiv.org/abs/1603.05737}{{\ttfamily
  arXiv:1603.05737 [hep-th]}}.

\bibitem{91}
W.~D. Goldberger and A.~K. Ridgway, ``{Radiation and the classical double copy
  for color charges},''
  \href{http://dx.doi.org/10.1103/PhysRevD.95.125010}{{\em Phys. Rev. D}
  {\bfseries 95} no.~12, (2017) 125010},
  \href{http://arxiv.org/abs/1611.03493}{{\ttfamily arXiv:1611.03493
  [hep-th]}}.

\bibitem{92}
A.~Luna, R.~Monteiro, I.~Nicholson, A.~Ochirov, D.~O'Connell, N.~Westerberg,
  and C.~D. White, ``{Perturbative spacetimes from Yang-Mills theory},''
  \href{http://dx.doi.org/10.1007/JHEP04(2017)069}{{\em JHEP} {\bfseries 04}
  (2017) 069}, \href{http://arxiv.org/abs/1611.07508}{{\ttfamily
  arXiv:1611.07508 [hep-th]}}.

\bibitem{96}
W.~D. Goldberger and A.~K. Ridgway, ``{Bound states and the classical double
  copy},'' \href{http://dx.doi.org/10.1103/PhysRevD.97.085019}{{\em Phys. Rev.
  D} {\bfseries 97} no.~8, (2018) 085019},
  \href{http://arxiv.org/abs/1711.09493}{{\ttfamily arXiv:1711.09493
  [hep-th]}}.

\bibitem{97}
J.~Plefka, J.~Steinhoff, and W.~Wormsbecher, ``{Effective action of dilaton
  gravity as the classical double copy of Yang-Mills theory},''
  \href{http://dx.doi.org/10.1103/PhysRevD.99.024021}{{\em Phys. Rev. D}
  {\bfseries 99} no.~2, (2019) 024021},
  \href{http://arxiv.org/abs/1807.09859}{{\ttfamily arXiv:1807.09859
  [hep-th]}}.

\bibitem{Johansson:2017srf}
H.~Johansson and J.~Nohle, ``{Conformal Gravity from Gauge Theory},''
  \href{http://arxiv.org/abs/1707.02965}{{\ttfamily arXiv:1707.02965
  [hep-th]}}.

\bibitem{Johansson:2018ues}
H.~Johansson, G.~Mogull, and F.~Teng, ``{Unraveling conformal gravity
  amplitudes},'' \href{http://dx.doi.org/10.1007/JHEP09(2018)080}{{\em JHEP}
  {\bfseries 09} (2018) 080}, \href{http://arxiv.org/abs/1806.05124}{{\ttfamily
  arXiv:1806.05124 [hep-th]}}.

\bibitem{Menezes:2021dyp}
G.~Menezes, ``{Color-kinematics duality, double copy and the unitarity method
  for higher-derivative QCD and quadratic gravity},''
  \href{http://dx.doi.org/10.1007/JHEP03(2022)074}{{\em JHEP} {\bfseries 03}
  (2022) 074}, \href{http://arxiv.org/abs/2112.00978}{{\ttfamily
  arXiv:2112.00978 [hep-th]}}.

\bibitem{Menezes:2022jow}
G.~Menezes, ``{Leading Singularities in Higher-Derivative
  Yang\textendash{}Mills Theory and Quadratic Gravity},''
  \href{http://dx.doi.org/10.3390/universe8060326}{{\em Universe} {\bfseries 8}
  no.~6, (2022) 326}, \href{http://arxiv.org/abs/2205.04996}{{\ttfamily
  arXiv:2205.04996 [hep-th]}}.

\bibitem{Lescano:2023pai}
E.~Lescano, G.~Menezes, and J.~A. Rodr\'\i{}guez, ``{Aspects of conformal
  gravity and double field theory from a double copy map},''
  \href{http://dx.doi.org/10.1103/PhysRevD.108.126017}{{\em Phys. Rev. D}
  {\bfseries 108} no.~12, (2023) 126017},
  \href{http://arxiv.org/abs/2307.14538}{{\ttfamily arXiv:2307.14538
  [hep-th]}}.

\bibitem{Azevedo:2018dgo}
T.~Azevedo, M.~Chiodaroli, H.~Johansson, and O.~Schlotterer, ``{Heterotic and
  bosonic string amplitudes via field theory},''
  \href{http://dx.doi.org/10.1007/JHEP10(2018)012}{{\em JHEP} {\bfseries 10}
  (2018) 012}, \href{http://arxiv.org/abs/1803.05452}{{\ttfamily
  arXiv:1803.05452 [hep-th]}}.

\bibitem{Azevedo:2019zbn}
T.~Azevedo, R.~L. Jusinskas, and M.~Lize, ``{Bosonic sectorized strings and the
  $(DF)^{2}$ theory},'' \href{http://dx.doi.org/10.1007/JHEP01(2020)082}{{\em
  JHEP} {\bfseries 01} (2020) 082},
  \href{http://arxiv.org/abs/1908.11371}{{\ttfamily arXiv:1908.11371
  [hep-th]}}.

\bibitem{LIGO1}
{\bfseries LIGO Scientific, Virgo} Collaboration, B.~P. Abbott {\em et al.},
  ``{Observation of Gravitational Waves from a Binary Black Hole Merger},''
  \href{http://dx.doi.org/10.1103/PhysRevLett.116.061102}{{\em Phys. Rev.
  Lett.} {\bfseries 116} no.~6, (2016) 061102},
  \href{http://arxiv.org/abs/1602.03837}{{\ttfamily arXiv:1602.03837 [gr-qc]}}.

\bibitem{LIGO2}
{\bfseries LIGO Scientific, Virgo} Collaboration, B.~P. Abbott {\em et al.},
  ``{GW151226: Observation of Gravitational Waves from a 22-Solar-Mass Binary
  Black Hole Coalescence},''
  \href{http://dx.doi.org/10.1103/PhysRevLett.116.241103}{{\em Phys. Rev.
  Lett.} {\bfseries 116} no.~24, (2016) 241103},
  \href{http://arxiv.org/abs/1606.04855}{{\ttfamily arXiv:1606.04855 [gr-qc]}}.

\bibitem{LIGO3}
{\bfseries LIGO Scientific, VIRGO} Collaboration, B.~P. Abbott {\em et al.},
  ``{GW170104: Observation of a 50-Solar-Mass Binary Black Hole Coalescence at
  Redshift 0.2},'' \href{http://dx.doi.org/10.1103/PhysRevLett.118.221101}{{\em
  Phys. Rev. Lett.} {\bfseries 118} no.~22, (2017) 221101},
  \href{http://arxiv.org/abs/1706.01812}{{\ttfamily arXiv:1706.01812 [gr-qc]}}.
  [Erratum: Phys.Rev.Lett. 121, 129901 (2018)].

\bibitem{LIGO4}
{\bfseries LIGO Scientific, Virgo} Collaboration, B.~P. Abbott {\em et al.},
  ``{GW170814: A Three-Detector Observation of Gravitational Waves from a
  Binary Black Hole Coalescence},''
  \href{http://dx.doi.org/10.1103/PhysRevLett.119.141101}{{\em Phys. Rev.
  Lett.} {\bfseries 119} no.~14, (2017) 141101},
  \href{http://arxiv.org/abs/1709.09660}{{\ttfamily arXiv:1709.09660 [gr-qc]}}.

\bibitem{LIGO5}
{\bfseries LIGO Scientific, Virgo} Collaboration, B.~P. Abbott {\em et al.},
  ``{GW170817: Observation of Gravitational Waves from a Binary Neutron Star
  Inspiral},'' \href{http://dx.doi.org/10.1103/PhysRevLett.119.161101}{{\em
  Phys. Rev. Lett.} {\bfseries 119} no.~16, (2017) 161101},
  \href{http://arxiv.org/abs/1710.05832}{{\ttfamily arXiv:1710.05832 [gr-qc]}}.

\bibitem{Bjerrum-Bohr:2022blt}
N.~E.~J. Bjerrum-Bohr, P.~H. Damgaard, L.~Plante, and P.~Vanhove, ``{The SAGEX
  review on scattering amplitudes Chapter 13: Post-Minkowskian expansion from
  scattering amplitudes},''
  \href{http://dx.doi.org/10.1088/1751-8121/ac7a78}{{\em J. Phys. A} {\bfseries
  55} no.~44, (2022) 443014}, \href{http://arxiv.org/abs/2203.13024}{{\ttfamily
  arXiv:2203.13024 [hep-th]}}.

\bibitem{Kosower:2022yvp}
D.~A. Kosower, R.~Monteiro, and D.~O'Connell, ``{The SAGEX review on scattering
  amplitudes Chapter 14: Classical gravity from scattering amplitudes},''
  \href{http://dx.doi.org/10.1088/1751-8121/ac8846}{{\em J. Phys. A} {\bfseries
  55} no.~44, (2022) 443015}, \href{http://arxiv.org/abs/2203.13025}{{\ttfamily
  arXiv:2203.13025 [hep-th]}}.

\bibitem{Kosower:19}
D.~A. Kosower, B.~Maybee, and D.~O'Connell, ``{Amplitudes, Observables, and
  Classical Scattering},''
  \href{http://dx.doi.org/10.1007/JHEP02(2019)137}{{\em JHEP} {\bfseries 02}
  (2019) 137}, \href{http://arxiv.org/abs/1811.10950}{{\ttfamily
  arXiv:1811.10950 [hep-th]}}.

\bibitem{Cachazo:2017jef}
F.~Cachazo and A.~Guevara, ``{Leading Singularities and Classical Gravitational
  Scattering},'' \href{http://dx.doi.org/10.1007/JHEP02(2020)181}{{\em JHEP}
  {\bfseries 02} (2020) 181}, \href{http://arxiv.org/abs/1705.10262}{{\ttfamily
  arXiv:1705.10262 [hep-th]}}.

\bibitem{Guevara:2017csg}
A.~Guevara, ``{Holomorphic Classical Limit for Spin Effects in Gravitational
  and Electromagnetic Scattering},''
  \href{http://dx.doi.org/10.1007/JHEP04(2019)033}{{\em JHEP} {\bfseries 04}
  (2019) 033}, \href{http://arxiv.org/abs/1706.02314}{{\ttfamily
  arXiv:1706.02314 [hep-th]}}.

\bibitem{Sturani:2021ucg}
R.~Sturani, ``{Fundamental Gravity and Gravitational Waves},''
  \href{http://dx.doi.org/10.3390/sym13122384}{{\em Symmetry} {\bfseries 13}
  no.~12, (2021) 2384}.

\bibitem{Neill:2013wsa}
D.~Neill and I.~Z. Rothstein, ``{Classical Space-Times from the S Matrix},''
  \href{http://dx.doi.org/10.1016/j.nuclphysb.2013.09.007}{{\em Nucl. Phys. B}
  {\bfseries 877} (2013) 177--189},
  \href{http://arxiv.org/abs/1304.7263}{{\ttfamily arXiv:1304.7263 [hep-th]}}.

\bibitem{Bjerrum-Bohr:14}
N.~E.~J. Bjerrum-Bohr, J.~F. Donoghue, and P.~Vanhove, ``{On-shell Techniques
  and Universal Results in Quantum Gravity},''
  \href{http://dx.doi.org/10.1007/JHEP02(2014)111}{{\em JHEP} {\bfseries 02}
  (2014) 111}, \href{http://arxiv.org/abs/1309.0804}{{\ttfamily arXiv:1309.0804
  [hep-th]}}.

\bibitem{Bjerrum-Bohr:2014lea}
N.~E.~J. Bjerrum-Bohr, B.~R. Holstein, L.~Plant\'e, and P.~Vanhove,
  ``{Graviton-Photon Scattering},''
  \href{http://dx.doi.org/10.1103/PhysRevD.91.064008}{{\em Phys. Rev. D}
  {\bfseries 91} no.~6, (2015) 064008},
  \href{http://arxiv.org/abs/1410.4148}{{\ttfamily arXiv:1410.4148 [gr-qc]}}.

\bibitem{Bjerrum-Bohr:2014zsa}
N.~E.~J. Bjerrum-Bohr, J.~F. Donoghue, B.~R. Holstein, L.~Plant\'e, and
  P.~Vanhove, ``{Bending of Light in Quantum Gravity},''
  \href{http://dx.doi.org/10.1103/PhysRevLett.114.061301}{{\em Phys. Rev.
  Lett.} {\bfseries 114} no.~6, (2015) 061301},
  \href{http://arxiv.org/abs/1410.7590}{{\ttfamily arXiv:1410.7590 [hep-th]}}.

\bibitem{Bjerrum-Bohr:16}
N.~E.~J. Bjerrum-Bohr, J.~F. Donoghue, B.~R. Holstein, L.~Plante, and
  P.~Vanhove, ``{Light-like Scattering in Quantum Gravity},''
  \href{http://dx.doi.org/10.1007/JHEP11(2016)117}{{\em JHEP} {\bfseries 11}
  (2016) 117}, \href{http://arxiv.org/abs/1609.07477}{{\ttfamily
  arXiv:1609.07477 [hep-th]}}.

\bibitem{76}
N.~E.~J. Bjerrum-Bohr, P.~H. Damgaard, G.~Festuccia, L.~Plant\'e, and
  P.~Vanhove, ``{General Relativity from Scattering Amplitudes},''
  \href{http://dx.doi.org/10.1103/PhysRevLett.121.171601}{{\em Phys. Rev.
  Lett.} {\bfseries 121} no.~17, (2018) 171601},
  \href{http://arxiv.org/abs/1806.04920}{{\ttfamily arXiv:1806.04920
  [hep-th]}}.

\bibitem{Bjerrum-Bohr:2021vuf}
N.~E.~J. Bjerrum-Bohr, P.~H. Damgaard, L.~Plant\'e, and P.~Vanhove,
  ``{Classical gravity from loop amplitudes},''
  \href{http://dx.doi.org/10.1103/PhysRevD.104.026009}{{\em Phys. Rev. D}
  {\bfseries 104} no.~2, (2021) 026009},
  \href{http://arxiv.org/abs/2104.04510}{{\ttfamily arXiv:2104.04510
  [hep-th]}}.

\bibitem{Bjerrum-Bohr:2021din}
N.~E.~J. Bjerrum-Bohr, P.~H. Damgaard, L.~Plant\'e, and P.~Vanhove, ``{The
  amplitude for classical gravitational scattering at third Post-Minkowskian
  order},'' \href{http://dx.doi.org/10.1007/JHEP08(2021)172}{{\em JHEP}
  {\bfseries 08} (2021) 172}, \href{http://arxiv.org/abs/2105.05218}{{\ttfamily
  arXiv:2105.05218 [hep-th]}}.

\bibitem{Herrmann:2021tct}
E.~Herrmann, J.~Parra-Martinez, M.~S. Ruf, and M.~Zeng, ``{Radiative classical
  gravitational observables at $ \mathcal{O} $(G$^{3}$) from scattering
  amplitudes},'' \href{http://dx.doi.org/10.1007/JHEP10(2021)148}{{\em JHEP}
  {\bfseries 10} (2021) 148}, \href{http://arxiv.org/abs/2104.03957}{{\ttfamily
  arXiv:2104.03957 [hep-th]}}.

\bibitem{Cristofoli:2021vyo}
A.~Cristofoli, R.~Gonzo, D.~A. Kosower, and D.~O'Connell, ``{Waveforms from
  amplitudes},'' \href{http://dx.doi.org/10.1103/PhysRevD.106.056007}{{\em
  Phys. Rev. D} {\bfseries 106} no.~5, (2022) 056007},
  \href{http://arxiv.org/abs/2107.10193}{{\ttfamily arXiv:2107.10193
  [hep-th]}}.

\bibitem{Bern:2021dqo}
Z.~Bern, J.~Parra-Martinez, R.~Roiban, M.~S. Ruf, C.-H. Shen, M.~P. Solon, and
  M.~Zeng, ``{Scattering Amplitudes and Conservative Binary Dynamics at ${\cal
  O}(G^4)$},'' \href{http://dx.doi.org/10.1103/PhysRevLett.126.171601}{{\em
  Phys. Rev. Lett.} {\bfseries 126} no.~17, (2021) 171601},
  \href{http://arxiv.org/abs/2101.07254}{{\ttfamily arXiv:2101.07254
  [hep-th]}}.

\bibitem{Bern:2021yeh}
Z.~Bern, J.~Parra-Martinez, R.~Roiban, M.~S. Ruf, C.-H. Shen, M.~P. Solon, and
  M.~Zeng, ``{Scattering Amplitudes, the Tail Effect, and Conservative Binary
  Dynamics at O(G4)},''
  \href{http://dx.doi.org/10.1103/PhysRevLett.128.161103}{{\em Phys. Rev.
  Lett.} {\bfseries 128} no.~16, (2022) 161103},
  \href{http://arxiv.org/abs/2112.10750}{{\ttfamily arXiv:2112.10750
  [hep-th]}}.

\bibitem{Herrmann:2021lqe}
E.~Herrmann, J.~Parra-Martinez, M.~S. Ruf, and M.~Zeng, ``{Gravitational
  Bremsstrahlung from Reverse Unitarity},''
  \href{http://dx.doi.org/10.1103/PhysRevLett.126.201602}{{\em Phys. Rev.
  Lett.} {\bfseries 126} no.~20, (2021) 201602},
  \href{http://arxiv.org/abs/2101.07255}{{\ttfamily arXiv:2101.07255
  [hep-th]}}.

\bibitem{DiVecchia:2021bdo}
P.~Di~Vecchia, C.~Heissenberg, R.~Russo, and G.~Veneziano, ``{The eikonal
  approach to gravitational scattering and radiation at $ \mathcal{O}
  $(G$^{3}$)},'' \href{http://dx.doi.org/10.1007/JHEP07(2021)169}{{\em JHEP}
  {\bfseries 07} (2021) 169}, \href{http://arxiv.org/abs/2104.03256}{{\ttfamily
  arXiv:2104.03256 [hep-th]}}.

\bibitem{Bern:2020gjj}
Z.~Bern, H.~Ita, J.~Parra-Martinez, and M.~S. Ruf, ``{Universality in the
  classical limit of massless gravitational scattering},''
  \href{http://dx.doi.org/10.1103/PhysRevLett.125.031601}{{\em Phys. Rev.
  Lett.} {\bfseries 125} no.~3, (2020) 031601},
  \href{http://arxiv.org/abs/2002.02459}{{\ttfamily arXiv:2002.02459
  [hep-th]}}.

\bibitem{Moynihan:2020gxj}
N.~Moynihan and J.~Murugan, ``{On-Shell Electric-Magnetic Duality and the Dual
  Graviton},'' \href{http://arxiv.org/abs/2002.11085}{{\ttfamily
  arXiv:2002.11085 [hep-th]}}.

\bibitem{Cristofoli:2020uzm}
A.~Cristofoli, P.~H. Damgaard, P.~Di~Vecchia, and C.~Heissenberg,
  ``{Second-order Post-Minkowskian scattering in arbitrary dimensions},''
  \href{http://dx.doi.org/10.1007/JHEP07(2020)122}{{\em JHEP} {\bfseries 07}
  (2020) 122}, \href{http://arxiv.org/abs/2003.10274}{{\ttfamily
  arXiv:2003.10274 [hep-th]}}.

\bibitem{Parra-Martinez:2020dzs}
J.~Parra-Martinez, M.~S. Ruf, and M.~Zeng, ``{Extremal black hole scattering at
  $\mathcal{O}(G^3)$: graviton dominance, eikonal exponentiation, and
  differential equations},''
  \href{http://dx.doi.org/10.1007/JHEP11(2020)023}{{\em JHEP} {\bfseries 11}
  (2020) 023}, \href{http://arxiv.org/abs/2005.04236}{{\ttfamily
  arXiv:2005.04236 [hep-th]}}.

\bibitem{Haddad:2020tvs}
K.~Haddad and A.~Helset, ``{The double copy for heavy particles},''
  \href{http://dx.doi.org/10.1103/PhysRevLett.125.181603}{{\em Phys. Rev.
  Lett.} {\bfseries 125} (2020) 181603},
  \href{http://arxiv.org/abs/2005.13897}{{\ttfamily arXiv:2005.13897
  [hep-th]}}.

\bibitem{AccettulliHuber:2020oou}
M.~Accettulli~Huber, A.~Brandhuber, S.~De~Angelis, and G.~Travaglini,
  ``{Eikonal phase matrix, deflection angle and time delay in effective field
  theories of gravity},''
  \href{http://dx.doi.org/10.1103/PhysRevD.102.046014}{{\em Phys. Rev. D}
  {\bfseries 102} no.~4, (2020) 046014},
  \href{http://arxiv.org/abs/2006.02375}{{\ttfamily arXiv:2006.02375
  [hep-th]}}.

\bibitem{Moynihan:2020ejh}
N.~Moynihan, ``{Scattering Amplitudes and the Double Copy in Topologically
  Massive Theories},'' \href{http://dx.doi.org/10.1007/JHEP12(2020)163}{{\em
  JHEP} {\bfseries 12} (2020) 163},
  \href{http://arxiv.org/abs/2006.15957}{{\ttfamily arXiv:2006.15957
  [hep-th]}}.

\bibitem{Manu:2020zxl}
A.~Manu, D.~Ghosh, A.~Laddha, and P.~V. Athira, ``{Soft radiation from
  scattering amplitudes revisited},''
  \href{http://dx.doi.org/10.1007/JHEP05(2021)056}{{\em JHEP} {\bfseries 05}
  (2021) 056}, \href{http://arxiv.org/abs/2007.02077}{{\ttfamily
  arXiv:2007.02077 [hep-th]}}.

\bibitem{Sahoo:2020ryf}
B.~Sahoo, ``{Classical Sub-subleading Soft Photon and Soft Graviton Theorems in
  Four Spacetime Dimensions},''
  \href{http://dx.doi.org/10.1007/JHEP12(2020)070}{{\em JHEP} {\bfseries 12}
  (2020) 070}, \href{http://arxiv.org/abs/2008.04376}{{\ttfamily
  arXiv:2008.04376 [hep-th]}}.

\bibitem{delaCruz:2020bbn}
L.~de~la Cruz, B.~Maybee, D.~O'Connell, and A.~Ross, ``{Classical Yang-Mills
  observables from amplitudes},''
  \href{http://dx.doi.org/10.1007/JHEP12(2020)076}{{\em JHEP} {\bfseries 12}
  (2020) 076}, \href{http://arxiv.org/abs/2009.03842}{{\ttfamily
  arXiv:2009.03842 [hep-th]}}.

\bibitem{Bonocore:2020xuj}
D.~Bonocore, ``{Asymptotic dynamics on the worldline for spinning particles},''
  \href{http://dx.doi.org/10.1007/JHEP02(2021)007}{{\em JHEP} {\bfseries 02}
  (2021) 007}, \href{http://arxiv.org/abs/2009.07863}{{\ttfamily
  arXiv:2009.07863 [hep-th]}}.

\bibitem{Mogull:2020sak}
G.~Mogull, J.~Plefka, and J.~Steinhoff, ``{Classical black hole scattering from
  a worldline quantum field theory},''
  \href{http://dx.doi.org/10.1007/JHEP02(2021)048}{{\em JHEP} {\bfseries 02}
  (2021) 048}, \href{http://arxiv.org/abs/2010.02865}{{\ttfamily
  arXiv:2010.02865 [hep-th]}}.

\bibitem{Emond:2020lwi}
W.~T. Emond, Y.-T. Huang, U.~Kol, N.~Moynihan, and D.~O'Connell, ``{Amplitudes
  from Coulomb to Kerr-Taub-NUT},''
  \href{http://arxiv.org/abs/2010.07861}{{\ttfamily arXiv:2010.07861
  [hep-th]}}.

\bibitem{Cheung:2020gbf}
C.~Cheung, N.~Shah, and M.~P. Solon, ``{Mining the Geodesic Equation for
  Scattering Data},'' \href{http://dx.doi.org/10.1103/PhysRevD.103.024030}{{\em
  Phys. Rev. D} {\bfseries 103} no.~2, (2021) 024030},
  \href{http://arxiv.org/abs/2010.08568}{{\ttfamily arXiv:2010.08568
  [hep-th]}}.

\bibitem{Mougiakakos:2020laz}
S.~Mougiakakos and P.~Vanhove, ``{Schwarzschild-Tangherlini metric from
  scattering amplitudes in various dimensions},''
  \href{http://dx.doi.org/10.1103/PhysRevD.103.026001}{{\em Phys. Rev. D}
  {\bfseries 103} no.~2, (2021) 026001},
  \href{http://arxiv.org/abs/2010.08882}{{\ttfamily arXiv:2010.08882
  [hep-th]}}.

\bibitem{Carrasco:2020ywq}
J.~J.~M. Carrasco and I.~A. Vazquez-Holm, ``{Loop-Level Double-Copy for Massive
  Quantum Particles},''
  \href{http://dx.doi.org/10.1103/PhysRevD.103.045002}{{\em Phys. Rev. D}
  {\bfseries 103} no.~4, (2021) 045002},
  \href{http://arxiv.org/abs/2010.13435}{{\ttfamily arXiv:2010.13435
  [hep-th]}}.

\bibitem{Kim:2020cvf}
J.-W. Kim and M.~Shim, ``{Gravitational Dyonic Amplitude at One-Loop and its
  Inconsistency with the Classical Impulse},''
  \href{http://dx.doi.org/10.1007/JHEP02(2021)217}{{\em JHEP} {\bfseries 02}
  (2021) 217}, \href{http://arxiv.org/abs/2010.14347}{{\ttfamily
  arXiv:2010.14347 [hep-th]}}.

\bibitem{Bjerrum-Bohr:2020syg}
N.~E.~J. Bjerrum-Bohr, T.~V. Brown, and H.~Gomez, ``{Scattering of Gravitons
  and Spinning Massive States from Compact Numerators},''
  \href{http://dx.doi.org/10.1007/JHEP04(2021)234}{{\em JHEP} {\bfseries 04}
  (2021) 234}, \href{http://arxiv.org/abs/2011.10556}{{\ttfamily
  arXiv:2011.10556 [hep-th]}}.

\bibitem{Gonzo:2020xza}
R.~Gonzo and A.~Pokraka, ``{Light-ray operators, detectors and gravitational
  event shapes},'' \href{http://dx.doi.org/10.1007/JHEP05(2021)015}{{\em JHEP}
  {\bfseries 05} (2021) 015}, \href{http://arxiv.org/abs/2012.01406}{{\ttfamily
  arXiv:2012.01406 [hep-th]}}.

\bibitem{delaCruz:2020cpc}
L.~de~la Cruz, ``{Scattering amplitudes approach to hard thermal loops},''
  \href{http://dx.doi.org/10.1103/PhysRevD.104.014013}{{\em Phys. Rev. D}
  {\bfseries 104} no.~1, (2021) 014013},
  \href{http://arxiv.org/abs/2012.07714}{{\ttfamily arXiv:2012.07714
  [hep-th]}}.

\bibitem{Cristofoli:2021jas}
A.~Cristofoli, R.~Gonzo, N.~Moynihan, D.~O'Connell, A.~Ross, M.~Sergola, and
  C.~D. White, ``{The Uncertainty Principle and Classical Amplitudes},''
  \href{http://arxiv.org/abs/2112.07556}{{\ttfamily arXiv:2112.07556
  [hep-th]}}.

\bibitem{Bautista:2021llr}
Y.~F. Bautista and A.~Laddha, ``{Soft Constraints on KMOC Formalism},''
  \href{http://arxiv.org/abs/2111.11642}{{\ttfamily arXiv:2111.11642
  [hep-th]}}.

\bibitem{Brandhuber:2021eyq}
A.~Brandhuber, G.~Chen, G.~Travaglini, and C.~Wen, ``{Classical gravitational
  scattering from a gauge-invariant double copy},''
  \href{http://dx.doi.org/10.1007/JHEP10(2021)118}{{\em JHEP} {\bfseries 10}
  (2021) 118}, \href{http://arxiv.org/abs/2108.04216}{{\ttfamily
  arXiv:2108.04216 [hep-th]}}.

\bibitem{Brandhuber:2021bsf}
A.~Brandhuber, G.~Chen, H.~Johansson, G.~Travaglini, and C.~Wen, ``{Kinematic
  Hopf Algebra for Bern-Carrasco-Johansson Numerators in Heavy-Mass Effective
  Field Theory and Yang-Mills Theory},''
  \href{http://dx.doi.org/10.1103/PhysRevLett.128.121601}{{\em Phys. Rev.
  Lett.} {\bfseries 128} no.~12, (2022) 121601},
  \href{http://arxiv.org/abs/2111.15649}{{\ttfamily arXiv:2111.15649
  [hep-th]}}.

\bibitem{Aoude:2021oqj}
R.~Aoude and A.~Ochirov, ``{Classical observables from coherent-spin
  amplitudes},'' \href{http://dx.doi.org/10.1007/JHEP10(2021)008}{{\em JHEP}
  {\bfseries 10} (2021) 008}, \href{http://arxiv.org/abs/2108.01649}{{\ttfamily
  arXiv:2108.01649 [hep-th]}}.

\bibitem{Cho:2022syn}
G.~Cho, R.~A. Porto, and Z.~Yang, ``{Gravitational radiation from inspiralling
  compact objects: Spin effects to the fourth post-Newtonian order},''
  \href{http://dx.doi.org/10.1103/PhysRevD.106.L101501}{{\em Phys. Rev. D}
  {\bfseries 106} no.~10, (2022) L101501},
  \href{http://arxiv.org/abs/2201.05138}{{\ttfamily arXiv:2201.05138 [gr-qc]}}.

\bibitem{Bern:2020buy}
Z.~Bern, A.~Luna, R.~Roiban, C.-H. Shen, and M.~Zeng, ``{Spinning black hole
  binary dynamics, scattering amplitudes, and effective field theory},''
  \href{http://dx.doi.org/10.1103/PhysRevD.104.065014}{{\em Phys. Rev. D}
  {\bfseries 104} no.~6, (2021) 065014},
  \href{http://arxiv.org/abs/2005.03071}{{\ttfamily arXiv:2005.03071
  [hep-th]}}.

\bibitem{Bautista:2021wfy}
Y.~F. Bautista, A.~Guevara, C.~Kavanagh, and J.~Vines, ``{Scattering in black
  hole backgrounds and higher-spin amplitudes. Part I},''
  \href{http://dx.doi.org/10.1007/JHEP03(2023)136}{{\em JHEP} {\bfseries 03}
  (2023) 136}, \href{http://arxiv.org/abs/2107.10179}{{\ttfamily
  arXiv:2107.10179 [hep-th]}}.

\bibitem{yutinspin}
W.-M. Chen, M.-Z. Chung, Y.-t. Huang, and J.-W. Kim, ``{The 2PM Hamiltonian for
  binary Kerr to quartic in spin},''
  \href{http://arxiv.org/abs/2111.13639}{{\ttfamily arXiv:2111.13639
  [hep-th]}}.

\bibitem{Alessio:2022kwv}
F.~Alessio and P.~Di~Vecchia, ``{Radiation reaction for spinning black-hole
  scattering},'' \href{http://arxiv.org/abs/2203.13272}{{\ttfamily
  arXiv:2203.13272 [hep-th]}}.

\bibitem{Bern:2022kto}
Z.~Bern, D.~Kosmopoulos, A.~Luna, R.~Roiban, and F.~Teng, ``{Binary Dynamics
  through the Fifth Power of Spin at O(G2)},''
  \href{http://dx.doi.org/10.1103/PhysRevLett.130.201402}{{\em Phys. Rev.
  Lett.} {\bfseries 130} no.~20, (2023) 201402},
  \href{http://arxiv.org/abs/2203.06202}{{\ttfamily arXiv:2203.06202
  [hep-th]}}.

\bibitem{FebresCordero:2022jts}
F.~Febres~Cordero, M.~Kraus, G.~Lin, M.~S. Ruf, and M.~Zeng, ``{Conservative
  Binary Dynamics with a Spinning Black Hole at O(G3) from Scattering
  Amplitudes},'' \href{http://dx.doi.org/10.1103/PhysRevLett.130.021601}{{\em
  Phys. Rev. Lett.} {\bfseries 130} no.~2, (2023) 021601},
  \href{http://arxiv.org/abs/2205.07357}{{\ttfamily arXiv:2205.07357
  [hep-th]}}.

\bibitem{Bohnenblust:2023qmy}
L.~Bohnenblust, H.~Ita, M.~Kraus, and J.~Schlenk, ``{Gravitational
  Bremsstrahlung in Black-Hole Scattering at $\mathcal{O}(G^3)$: Linear-in-Spin
  Effects},'' \href{http://arxiv.org/abs/2312.14859}{{\ttfamily
  arXiv:2312.14859 [hep-th]}}.

\bibitem{Maybee:2019jus}
B.~Maybee, D.~O'Connell, and J.~Vines, ``{Observables and amplitudes for
  spinning particles and black holes},''
  \href{http://dx.doi.org/10.1007/JHEP12(2019)156}{{\em JHEP} {\bfseries 12}
  (2019) 156}, \href{http://arxiv.org/abs/1906.09260}{{\ttfamily
  arXiv:1906.09260 [hep-th]}}.

\bibitem{Foffa:2019yfl}
S.~Foffa, R.~A. Porto, I.~Rothstein, and R.~Sturani, ``{Conservative dynamics
  of binary systems to fourth Post-Newtonian order in the EFT approach II:
  Renormalized Lagrangian},''
  \href{http://dx.doi.org/10.1103/PhysRevD.100.024048}{{\em Phys. Rev.}
  {\bfseries D100} no.~2, (2019) 024048},
\href{http://arxiv.org/abs/1903.05118}{{\ttfamily arXiv:1903.05118 [gr-qc]}}.

\bibitem{Bern:2019nnu}
Z.~Bern, C.~Cheung, R.~Roiban, C.-H. Shen, M.~P. Solon, and M.~Zeng,
  ``{Scattering Amplitudes and the Conservative Hamiltonian for Binary Systems
  at Third Post-Minkowskian Order},''
  \href{http://dx.doi.org/10.1103/PhysRevLett.122.201603}{{\em Phys. Rev.
  Lett.} {\bfseries 122} no.~20, (2019) 201603},
\href{http://arxiv.org/abs/1901.04424}{{\ttfamily arXiv:1901.04424 [hep-th]}}.

\bibitem{Bini:2019nra}
D.~Bini, T.~Damour, and A.~Geralico, ``{Novel approach to binary dynamics:
  application to the fifth post-Newtonian level},''
  \href{http://dx.doi.org/10.1103/PhysRevLett.123.231104}{{\em Phys. Rev.
  Lett.} {\bfseries 123} no.~23, (2019) 231104},
  \href{http://arxiv.org/abs/1909.02375}{{\ttfamily arXiv:1909.02375 [gr-qc]}}.

\bibitem{Bini:2020wpo}
D.~Bini, T.~Damour, and A.~Geralico, ``{Binary dynamics at the fifth and
  fifth-and-a-half post-Newtonian orders},''
  \href{http://dx.doi.org/10.1103/PhysRevD.102.024062}{{\em Phys. Rev. D}
  {\bfseries 102} no.~2, (2020) 024062},
  \href{http://arxiv.org/abs/2003.11891}{{\ttfamily arXiv:2003.11891 [gr-qc]}}.

\bibitem{18}
W.~D. Goldberger and I.~Z. Rothstein, ``{An Effective field theory of gravity
  for extended objects},''
  \href{http://dx.doi.org/10.1103/PhysRevD.73.104029}{{\em Phys. Rev. D}
  {\bfseries 73} (2006) 104029},
  \href{http://arxiv.org/abs/hep-th/0409156}{{\ttfamily arXiv:hep-th/0409156}}.

\bibitem{19}
R.~A. Porto, ``{The effective field theorist\textquoteright{}s approach to
  gravitational dynamics},''
  \href{http://dx.doi.org/10.1016/j.physrep.2016.04.003}{{\em Phys. Rept.}
  {\bfseries 633} (2016) 1--104},
  \href{http://arxiv.org/abs/1601.04914}{{\ttfamily arXiv:1601.04914
  [hep-th]}}.

\bibitem{Iwasaki:1971vb}
Y.~Iwasaki, ``{Quantum theory of gravitation vs. classical theory. -
  fourth-order potential},'' \href{http://dx.doi.org/10.1143/PTP.46.1587}{{\em
  Prog. Theor. Phys.} {\bfseries 46} (1971) 1587--1609}.

\bibitem{HariDass:1980tq}
N.~Hari~Dass and V.~Soni, ``{Feynman graph derivation of Einstein quadrupole
  formula},''
\href{http://dx.doi.org/10.1088/0305-4470/15/2/019}{{\em J.Phys.} {\bfseries
  A15} (1982) 473}.

\bibitem{Damour:1995kt}
T.~Damour and G.~Esposito-Farese, ``{Testing gravity to second postNewtonian
  order: A Field theory approach},''
  \href{http://dx.doi.org/10.1103/PhysRevD.53.5541}{{\em Phys.Rev.} {\bfseries
  D53} (1996) 5541--5578},
\href{http://arxiv.org/abs/gr-qc/9506063}{{\ttfamily arXiv:gr-qc/9506063
  [gr-qc]}}.

\bibitem{77}
J.~F. Donoghue, ``{Leading quantum correction to the Newtonian potential},''
  \href{http://dx.doi.org/10.1103/PhysRevLett.72.2996}{{\em Phys. Rev. Lett.}
  {\bfseries 72} (1994) 2996--2999},
  \href{http://arxiv.org/abs/gr-qc/9310024}{{\ttfamily arXiv:gr-qc/9310024}}.

\bibitem{78}
J.~F. Donoghue, ``{General relativity as an effective field theory: The leading
  quantum corrections},''
  \href{http://dx.doi.org/10.1103/PhysRevD.50.3874}{{\em Phys. Rev. D}
  {\bfseries 50} (1994) 3874--3888},
  \href{http://arxiv.org/abs/gr-qc/9405057}{{\ttfamily arXiv:gr-qc/9405057}}.

\bibitem{Kalin:2020mvi}
G.~K\"alin and R.~A. Porto, ``{Post-Minkowskian Effective Field Theory for
  Conservative Binary Dynamics},''
  \href{http://dx.doi.org/10.1007/JHEP11(2020)106}{{\em JHEP} {\bfseries 11}
  (2020) 106}, \href{http://arxiv.org/abs/2006.01184}{{\ttfamily
  arXiv:2006.01184 [hep-th]}}.

\bibitem{Kalin:2020fhe}
G.~K\"alin, Z.~Liu, and R.~A. Porto, ``{Conservative Dynamics of Binary Systems
  to Third Post-Minkowskian Order from the Effective Field Theory Approach},''
  \href{http://dx.doi.org/10.1103/PhysRevLett.125.261103}{{\em Phys. Rev.
  Lett.} {\bfseries 125} no.~26, (2020) 261103},
  \href{http://arxiv.org/abs/2007.04977}{{\ttfamily arXiv:2007.04977
  [hep-th]}}.

\bibitem{Foffa:2021pkg}
S.~Foffa and R.~Sturani, ``{Near and far zones in two-body dynamics: An
  effective field theory perspective},''
  \href{http://dx.doi.org/10.1103/PhysRevD.104.024069}{{\em Phys. Rev. D}
  {\bfseries 104} no.~2, (2021) 024069},
  \href{http://arxiv.org/abs/2103.03190}{{\ttfamily arXiv:2103.03190 [gr-qc]}}.

\bibitem{Almeida:2021xwn}
G.~L. Almeida, S.~Foffa, and R.~Sturani, ``{Tail contributions to gravitational
  conservative dynamics},''
  \href{http://dx.doi.org/10.1103/PhysRevD.104.124075}{{\em Phys. Rev. D}
  {\bfseries 104} no.~12, (2021) 124075},
  \href{http://arxiv.org/abs/2110.14146}{{\ttfamily arXiv:2110.14146 [gr-qc]}}.

\bibitem{Dlapa:2024cje}
C.~Dlapa, G.~K\"alin, Z.~Liu, and R.~A. Porto, ``{Local in Time Conservative
  Binary Dynamics at Fourth Post-Minkowskian Order},''
  \href{http://dx.doi.org/10.1103/PhysRevLett.132.221401}{{\em Phys. Rev.
  Lett.} {\bfseries 132} no.~22, (2024) 221401},
  \href{http://arxiv.org/abs/2403.04853}{{\ttfamily arXiv:2403.04853
  [hep-th]}}.

\bibitem{Foffa:2016rgu}
S.~Foffa, P.~Mastrolia, R.~Sturani, and C.~Sturm, ``{Effective field theory
  approach to the gravitational two-body dynamics, at fourth post-Newtonian
  order and quintic in the Newton constant},''
  \href{http://dx.doi.org/10.1103/PhysRevD.95.104009}{{\em Phys. Rev.}
  {\bfseries D95} no.~10, (2017) 104009},
\href{http://arxiv.org/abs/1612.00482}{{\ttfamily arXiv:1612.00482 [gr-qc]}}.

\bibitem{worldsheetpaper}
A.~Guevara, B.~Maybee, A.~Ochirov, D.~O'connell, and J.~Vines, ``{A worldsheet
  for Kerr},'' \href{http://dx.doi.org/10.1007/JHEP03(2021)201}{{\em JHEP}
  {\bfseries 03} (2021) 201}, \href{http://arxiv.org/abs/2012.11570}{{\ttfamily
  arXiv:2012.11570 [hep-th]}}.

\bibitem{Li:2018qap}
J.~Li and S.~G. Prabhu, ``{Gravitational radiation from the classical spinning
  double copy},'' \href{http://dx.doi.org/10.1103/PhysRevD.97.105019}{{\em
  Phys. Rev. D} {\bfseries 97} no.~10, (2018) 105019},
  \href{http://arxiv.org/abs/1803.02405}{{\ttfamily arXiv:1803.02405
  [hep-th]}}.

\bibitem{Goldberger:2017ogt}
W.~D. Goldberger, J.~Li, and S.~G. Prabhu, ``{Spinning particles, axion
  radiation, and the classical double copy},''
  \href{http://dx.doi.org/10.1103/PhysRevD.97.105018}{{\em Phys. Rev. D}
  {\bfseries 97} no.~10, (2018) 105018},
  \href{http://arxiv.org/abs/1712.09250}{{\ttfamily arXiv:1712.09250
  [hep-th]}}.

\bibitem{Bautista:2019evw}
Y.~F. Bautista and A.~Guevara, ``{On the Double Copy for Spinning Matter},''
  \href{http://arxiv.org/abs/1908.11349}{{\ttfamily arXiv:1908.11349
  [hep-th]}}.

\bibitem{Bautista:2022wjf}
Y.~F. Bautista, A.~Guevara, C.~Kavanagh, and J.~Vines, ``{Scattering in black
  hole backgrounds and higher-spin amplitudes. Part II},''
  \href{http://dx.doi.org/10.1007/JHEP05(2023)211}{{\em JHEP} {\bfseries 05}
  (2023) 211}, \href{http://arxiv.org/abs/2212.07965}{{\ttfamily
  arXiv:2212.07965 [hep-th]}}.

\bibitem{Menezes:2022tcs}
G.~Menezes and M.~Sergola, ``{NLO deflections for spinning particles and Kerr
  black holes},'' \href{http://dx.doi.org/10.1007/JHEP10(2022)105}{{\em JHEP}
  {\bfseries 10} (2022) 105}, \href{http://arxiv.org/abs/2205.11701}{{\ttfamily
  arXiv:2205.11701 [hep-th]}}.

\bibitem{Aoude:2022trd}
R.~Aoude, K.~Haddad, and A.~Helset, ``{Searching for Kerr in the 2PM
  amplitude},'' \href{http://dx.doi.org/10.1007/JHEP07(2022)072}{{\em JHEP}
  {\bfseries 07} (2022) 072}, \href{http://arxiv.org/abs/2203.06197}{{\ttfamily
  arXiv:2203.06197 [hep-th]}}.

\bibitem{Haddad:2023ylx}
K.~Haddad, ``{Recursion in the classical limit and the neutron-star Compton
  amplitude},'' \href{http://arxiv.org/abs/2303.02624}{{\ttfamily
  arXiv:2303.02624 [hep-th]}}.

\bibitem{Chiodaroli:2021eug}
M.~Chiodaroli, H.~Johansson, and P.~Pichini, ``{Compton black-hole scattering
  for s \ensuremath{\leq} 5/2},''
  \href{http://dx.doi.org/10.1007/JHEP02(2022)156}{{\em JHEP} {\bfseries 02}
  (2022) 156}, \href{http://arxiv.org/abs/2107.14779}{{\ttfamily
  arXiv:2107.14779 [hep-th]}}.

\bibitem{Chung:19}
M.-Z. Chung, Y.-T. Huang, J.-W. Kim, and S.~Lee, ``{The simplest massive
  S-matrix: from minimal coupling to Black Holes},''
  \href{http://dx.doi.org/10.1007/JHEP04(2019)156}{{\em JHEP} {\bfseries 04}
  (2019) 156}, \href{http://arxiv.org/abs/1812.08752}{{\ttfamily
  arXiv:1812.08752 [hep-th]}}.

\bibitem{Cangemi:2023ysz}
L.~Cangemi, M.~Chiodaroli, H.~Johansson, A.~Ochirov, P.~Pichini, and
  E.~Skvortsov, ``{From higher-spin gauge interactions to Compton amplitudes
  for root-Kerr},'' \href{http://arxiv.org/abs/2311.14668}{{\ttfamily
  arXiv:2311.14668 [hep-th]}}.

\bibitem{Cangemi:2023bpe}
L.~Cangemi, M.~Chiodaroli, H.~Johansson, A.~Ochirov, P.~Pichini, and
  E.~Skvortsov, ``{Compton Amplitude for Rotating Black Hole from QFT},''
  \href{http://dx.doi.org/10.1103/PhysRevLett.133.071601}{{\em Phys. Rev.
  Lett.} {\bfseries 133} no.~7, (2024) 071601},
  \href{http://arxiv.org/abs/2312.14913}{{\ttfamily arXiv:2312.14913
  [hep-th]}}.

\bibitem{Bjerrum-Bohr:2023jau}
N.~E.~J. Bjerrum-Bohr, G.~Chen, and M.~Skowronek, ``{Classical spin
  gravitational Compton scattering},''
  \href{http://dx.doi.org/10.1007/JHEP06(2023)170}{{\em JHEP} {\bfseries 06}
  (2023) 170}, \href{http://arxiv.org/abs/2302.00498}{{\ttfamily
  arXiv:2302.00498 [hep-th]}}.

\bibitem{Bjerrum-Bohr:2023iey}
N.~E.~J. Bjerrum-Bohr, G.~Chen, and M.~Skowronek, ``{Covariant Compton
  Amplitudes in Gravity with Classical Spin},''
  \href{http://dx.doi.org/10.1103/PhysRevLett.132.191603}{{\em Phys. Rev.
  Lett.} {\bfseries 132} no.~19, (2024) 191603},
  \href{http://arxiv.org/abs/2309.11249}{{\ttfamily arXiv:2309.11249
  [hep-th]}}.

\bibitem{yutinspin2}
M.-Z. Chung, Y.-t. Huang, J.-W. Kim, and S.~Lee, ``{Complete Hamiltonian for
  spinning binary systems at first post-Minkowskian order},''
  \href{http://dx.doi.org/10.1007/JHEP05(2020)105}{{\em JHEP} {\bfseries 05}
  (2020) 105}, \href{http://arxiv.org/abs/2003.06600}{{\ttfamily
  arXiv:2003.06600 [hep-th]}}.

\bibitem{mogullspin2}
G.~U. Jakobsen and G.~Mogull, ``{Conservative and Radiative Dynamics of
  Spinning Bodies at Third Post-Minkowskian Order Using Worldline Quantum Field
  Theory},'' \href{http://dx.doi.org/10.1103/PhysRevLett.128.141102}{{\em Phys.
  Rev. Lett.} {\bfseries 128} no.~14, (2022) 141102},
  \href{http://arxiv.org/abs/2201.07778}{{\ttfamily arXiv:2201.07778
  [hep-th]}}.

\bibitem{Guevara:2019fsj}
A.~Guevara, A.~Ochirov, and J.~Vines, ``{Black-hole scattering with general
  spin directions from minimal-coupling amplitudes},''
  \href{http://dx.doi.org/10.1103/PhysRevD.100.104024}{{\em Phys. Rev. D}
  {\bfseries 100} no.~10, (2019) 104024},
  \href{http://arxiv.org/abs/1906.10071}{{\ttfamily arXiv:1906.10071
  [hep-th]}}.

\bibitem{Guevara:19a}
A.~Guevara, A.~Ochirov, and J.~Vines, ``{Scattering of Spinning Black Holes
  from Exponentiated Soft Factors},''
  \href{http://dx.doi.org/10.1007/JHEP09(2019)056}{{\em JHEP} {\bfseries 09}
  (2019) 056}, \href{http://arxiv.org/abs/1812.06895}{{\ttfamily
  arXiv:1812.06895 [hep-th]}}.

\bibitem{Arkani-Hamed:20}
N.~Arkani-Hamed, Y.-t. Huang, and D.~O'Connell, ``{Kerr black holes as
  elementary particles},''
  \href{http://dx.doi.org/10.1007/JHEP01(2020)046}{{\em JHEP} {\bfseries 01}
  (2020) 046}, \href{http://arxiv.org/abs/1906.10100}{{\ttfamily
  arXiv:1906.10100 [hep-th]}}.

\bibitem{Damour:2024mzo}
T.~Damour, ``{Editorial note to Jean-Marie Souriau's '' On the motion of
  spinning particles in general relativity''},''
  \href{http://arxiv.org/abs/2401.10013}{{\ttfamily arXiv:2401.10013 [gr-qc]}}.

\bibitem{Bern:2023ity}
Z.~Bern, D.~Kosmopoulos, A.~Luna, R.~Roiban, T.~Scheopner, F.~Teng, and
  J.~Vines, ``{Quantum Field Theory, Worldline Theory, and Spin Magnitude
  Change in Orbital Evolution},''
  \href{http://arxiv.org/abs/2308.14176}{{\ttfamily arXiv:2308.14176
  [hep-th]}}.

\bibitem{Burger:2019wkq}
D.~J. Burger, W.~T. Emond, and N.~Moynihan, ``{Rotating Black Holes in Cubic
  Gravity},'' \href{http://dx.doi.org/10.1103/PhysRevD.101.084009}{{\em Phys.
  Rev. D} {\bfseries 101} no.~8, (2020) 084009},
  \href{http://arxiv.org/abs/1910.11618}{{\ttfamily arXiv:1910.11618
  [hep-th]}}.

\bibitem{Levi:2020lfn}
M.~Levi and F.~Teng, ``{NLO gravitational quartic-in-spin interaction},''
  \href{http://dx.doi.org/10.1007/JHEP01(2021)066}{{\em JHEP} {\bfseries 01}
  (2021) 066}, \href{http://arxiv.org/abs/2008.12280}{{\ttfamily
  arXiv:2008.12280 [hep-th]}}.

\bibitem{Levi:2020uwu}
M.~Levi, A.~J. Mcleod, and M.~Von~Hippel, ``{NNNLO gravitational
  quadratic-in-spin interactions at the quartic order in G},''
  \href{http://arxiv.org/abs/2003.07890}{{\ttfamily arXiv:2003.07890
  [hep-th]}}.

\bibitem{Liu:2021zxr}
Z.~Liu, R.~A. Porto, and Z.~Yang, ``{Spin Effects in the Effective Field Theory
  Approach to Post-Minkowskian Conservative Dynamics},''
  \href{http://dx.doi.org/10.1007/JHEP06(2021)012}{{\em JHEP} {\bfseries 06}
  (2021) 012}, \href{http://arxiv.org/abs/2102.10059}{{\ttfamily
  arXiv:2102.10059 [hep-th]}}.

\bibitem{Faye:2006gx}
G.~Faye, L.~Blanchet, and A.~Buonanno, ``{Higher-order spin effects in the
  dynamics of compact binaries. I. Equations of motion},''
  \href{http://dx.doi.org/10.1103/PhysRevD.74.104033}{{\em Phys. Rev. D}
  {\bfseries 74} (2006) 104033},
  \href{http://arxiv.org/abs/gr-qc/0605139}{{\ttfamily arXiv:gr-qc/0605139}}.

\bibitem{Blanchet:2006gy}
L.~Blanchet, A.~Buonanno, and G.~Faye, ``{Higher-order spin effects in the
  dynamics of compact binaries. II. Radiation field},''
  \href{http://dx.doi.org/10.1103/PhysRevD.81.089901}{{\em Phys. Rev. D}
  {\bfseries 74} (2006) 104034},
  \href{http://arxiv.org/abs/gr-qc/0605140}{{\ttfamily arXiv:gr-qc/0605140}}.
  [Erratum: Phys.Rev.D 75, 049903 (2007), Erratum: Phys.Rev.D 81, 089901
  (2010)].

\bibitem{Damour:2007nc}
T.~Damour, P.~Jaranowski, and G.~Schaefer, ``{Hamiltonian of two spinning
  compact bodies with next-to-leading order gravitational spin-orbit
  coupling},'' \href{http://dx.doi.org/10.1103/PhysRevD.77.064032}{{\em Phys.
  Rev. D} {\bfseries 77} (2008) 064032},
  \href{http://arxiv.org/abs/0711.1048}{{\ttfamily arXiv:0711.1048 [gr-qc]}}.

\bibitem{Vines:2017hyw}
J.~Vines, ``{Scattering of two spinning black holes in post-Minkowskian
  gravity, to all orders in spin, and effective-one-body mappings},''
  \href{http://dx.doi.org/10.1088/1361-6382/aaa3a8}{{\em Class. Quant. Grav.}
  {\bfseries 35} no.~8, (2018) 084002},
\href{http://arxiv.org/abs/1709.06016}{{\ttfamily arXiv:1709.06016 [gr-qc]}}.

\bibitem{Bini:2017xzy}
D.~Bini and T.~Damour, ``{Gravitational spin-orbit coupling in binary systems,
  post-Minkowskian approximation and effective one-body theory},''
  \href{http://dx.doi.org/10.1103/PhysRevD.96.104038}{{\em Phys. Rev. D}
  {\bfseries 96} no.~10, (2017) 104038},
  \href{http://arxiv.org/abs/1709.00590}{{\ttfamily arXiv:1709.00590 [gr-qc]}}.

\bibitem{Bini:2018ywr}
D.~Bini and T.~Damour, ``{Gravitational spin-orbit coupling in binary systems
  at the second post-Minkowskian approximation},''
  \href{http://dx.doi.org/10.1103/PhysRevD.98.044036}{{\em Phys. Rev. D}
  {\bfseries 98} no.~4, (2018) 044036},
  \href{http://arxiv.org/abs/1805.10809}{{\ttfamily arXiv:1805.10809 [gr-qc]}}.

\bibitem{Scheopner:2023rzp}
T.~Scheopner and J.~Vines, ``{Dynamical Implications of the Kerr Multipole
  Moments for Spinning Black Holes},''
  \href{http://arxiv.org/abs/2311.18421}{{\ttfamily arXiv:2311.18421 [gr-qc]}}.

\bibitem{Brandhuber:2023hhl}
A.~Brandhuber, G.~R. Brown, G.~Chen, J.~Gowdy, and G.~Travaglini, ``{Resummed
  spinning waveforms from five-point amplitudes},''
  \href{http://arxiv.org/abs/2310.04405}{{\ttfamily arXiv:2310.04405
  [hep-th]}}.

\bibitem{DeAngelis:2023lvf}
S.~De~Angelis, R.~Gonzo, and P.~P. Novichkov, ``{Spinning waveforms from KMOC
  at leading order},'' \href{http://arxiv.org/abs/2309.17429}{{\ttfamily
  arXiv:2309.17429 [hep-th]}}.

\bibitem{Mathisson:1937zz}
M.~Mathisson, ``{Neue mechanik materieller systemes},'' {\em Acta Phys. Polon.}
  {\bfseries 6} (1937) 163--200.

\bibitem{Papapetrou:1951pa}
A.~Papapetrou, ``{Spinning test particles in general relativity. 1.},''
  \href{http://dx.doi.org/10.1098/rspa.1951.0200}{{\em Proc. Roy. Soc. Lond. A}
  {\bfseries 209} (1951) 248--258}.

\bibitem{Pirani:1956tn}
F.~A.~E. Pirani, ``{On the Physical significance of the Riemann tensor},''
  \href{http://dx.doi.org/10.1007/s10714-009-0787-9}{{\em Acta Phys. Polon.}
  {\bfseries 15} (1956) 389--405}.

\bibitem{Tulczyjew:59}
W.~Tulczyjew, ``{Equations of Motion of Rotating Bodies in General Relativity
  Theory},'' {\em Acta Phys. Polon.} {\bfseries 18} (1959) 37--55.

\bibitem{Jakobsen:2021lvp}
G.~U. Jakobsen, G.~Mogull, J.~Plefka, and J.~Steinhoff, ``{Gravitational
  Bremsstrahlung and Hidden Supersymmetry of Spinning Bodies},''
  \href{http://dx.doi.org/10.1103/PhysRevLett.128.011101}{{\em Phys. Rev.
  Lett.} {\bfseries 128} no.~1, (2022) 011101},
  \href{http://arxiv.org/abs/2106.10256}{{\ttfamily arXiv:2106.10256
  [hep-th]}}.

\bibitem{Jakobsen:2022fcj}
G.~U. Jakobsen and G.~Mogull, ``{Conservative and Radiative Dynamics of
  Spinning Bodies at Third Post-Minkowskian Order Using Worldline Quantum Field
  Theory},'' \href{http://dx.doi.org/10.1103/PhysRevLett.128.141102}{{\em Phys.
  Rev. Lett.} {\bfseries 128} no.~14, (2022) 141102},
  \href{http://arxiv.org/abs/2201.07778}{{\ttfamily arXiv:2201.07778
  [hep-th]}}.

\bibitem{Riva:2022fru}
M.~M. Riva, F.~Vernizzi, and L.~K. Wong, ``{Gravitational bremsstrahlung from
  spinning binaries in the post-Minkowskian expansion},''
  \href{http://dx.doi.org/10.1103/PhysRevD.106.044013}{{\em Phys. Rev. D}
  {\bfseries 106} no.~4, (2022) 044013},
  \href{http://arxiv.org/abs/2205.15295}{{\ttfamily arXiv:2205.15295
  [hep-th]}}.

\bibitem{Aoude:2022thd}
R.~Aoude, K.~Haddad, and A.~Helset, ``{Classical gravitational
  spinning-spinless scattering at $\mathcal{O}(G^{2} S^{\infty})$},''
  \href{http://arxiv.org/abs/2205.02809}{{\ttfamily arXiv:2205.02809
  [hep-th]}}.

\bibitem{Aoude:2023vdk}
R.~Aoude, K.~Haddad, and A.~Helset, ``{Classical gravitational scattering
  amplitude at ${\cal O}(G^2
  S_1^{\ensuremath{\infty}}S_2^{\ensuremath{\infty}})$},''
  \href{http://dx.doi.org/10.1103/PhysRevD.108.024050}{{\em Phys. Rev. D}
  {\bfseries 108} no.~2, (2023) 024050},
  \href{http://arxiv.org/abs/2304.13740}{{\ttfamily arXiv:2304.13740
  [hep-th]}}.

\bibitem{Bautista:2021inx}
Y.~F. Bautista and N.~Siemonsen, ``{Post-Newtonian waveforms from spinning
  scattering amplitudes},''
  \href{http://dx.doi.org/10.1007/JHEP01(2022)006}{{\em JHEP} {\bfseries 01}
  (2022) 006}, \href{http://arxiv.org/abs/2110.12537}{{\ttfamily
  arXiv:2110.12537 [hep-th]}}.

\bibitem{Aoude:2023dui}
R.~Aoude, K.~Haddad, C.~Heissenberg, and A.~Helset, ``{Leading-order
  gravitational radiation to all spin orders},''
  \href{http://dx.doi.org/10.1103/PhysRevD.109.036007}{{\em Phys. Rev. D}
  {\bfseries 109} no.~3, (2024) 036007},
  \href{http://arxiv.org/abs/2310.05832}{{\ttfamily arXiv:2310.05832
  [hep-th]}}.

\bibitem{Arkani-Hamed:2017jhn}
N.~Arkani-Hamed, T.-C. Huang, and Y.-t. Huang, ``{Scattering Amplitudes For All
  Masses and Spins},''
\href{http://arxiv.org/abs/1709.04891}{{\ttfamily arXiv:1709.04891 [hep-th]}}.

\bibitem{Johansson:19}
H.~Johansson and A.~Ochirov, ``{Double copy for massive quantum particles with
  spin},'' \href{http://dx.doi.org/10.1007/JHEP09(2019)040}{{\em JHEP}
  {\bfseries 09} (2019) 040}, \href{http://arxiv.org/abs/1906.12292}{{\ttfamily
  arXiv:1906.12292 [hep-th]}}.

\bibitem{Skvortsov:2023jbn}
E.~Skvortsov and M.~Tsulaia, ``{Cubic action for spinning black holes from
  massive higher-spin gauge symmetry},''
  \href{http://dx.doi.org/10.1007/JHEP02(2024)202}{{\em JHEP} {\bfseries 02}
  (2024) 202}, \href{http://arxiv.org/abs/2312.08184}{{\ttfamily
  arXiv:2312.08184 [hep-th]}}.

\bibitem{Cangemi:2022bew}
L.~Cangemi, M.~Chiodaroli, H.~Johansson, A.~Ochirov, P.~Pichini, and
  E.~Skvortsov, ``{Kerr Black Holes From Massive Higher-Spin Gauge Symmetry},''
  \href{http://dx.doi.org/10.1103/PhysRevLett.131.221401}{{\em Phys. Rev.
  Lett.} {\bfseries 131} no.~22, (2023) 221401},
  \href{http://arxiv.org/abs/2212.06120}{{\ttfamily arXiv:2212.06120
  [hep-th]}}.

\bibitem{Schlotterer:2010kk}
O.~Schlotterer, ``{Higher Spin Scattering in Superstring Theory},''
  \href{http://dx.doi.org/10.1016/j.nuclphysb.2011.03.026}{{\em Nucl. Phys. B}
  {\bfseries 849} (2011) 433--460},
  \href{http://arxiv.org/abs/1011.1235}{{\ttfamily arXiv:1011.1235 [hep-th]}}.

\bibitem{Cangemi:2022abk}
L.~Cangemi and P.~Pichini, ``{Classical limit of higher-spin string
  amplitudes},'' \href{http://dx.doi.org/10.1007/JHEP06(2023)167}{{\em JHEP}
  {\bfseries 06} (2023) 167}, \href{http://arxiv.org/abs/2207.03947}{{\ttfamily
  arXiv:2207.03947 [hep-th]}}.

\bibitem{Giannakis:1998wi}
I.~Giannakis, J.~T. Liu, and M.~Porrati, ``{Massive higher spin states in
  string theory and the principle of equivalence},''
  \href{http://dx.doi.org/10.1103/PhysRevD.59.104013}{{\em Phys. Rev. D}
  {\bfseries 59} (1999) 104013},
  \href{http://arxiv.org/abs/hep-th/9809142}{{\ttfamily arXiv:hep-th/9809142}}.

\bibitem{Sagnotti:2010at}
A.~Sagnotti and M.~Taronna, ``{String Lessons for Higher-Spin Interactions},''
  \href{http://dx.doi.org/10.1016/j.nuclphysb.2010.08.019}{{\em Nucl. Phys. B}
  {\bfseries 842} (2011) 299--361},
  \href{http://arxiv.org/abs/1006.5242}{{\ttfamily arXiv:1006.5242 [hep-th]}}.

\bibitem{Agia:2023lfl}
N.~Agia, ``{Massive Type IIB Superstrings Part I: 3- and 4-Point Amplitudes},''
  \href{http://arxiv.org/abs/2309.11538}{{\ttfamily arXiv:2309.11538
  [hep-th]}}.

\bibitem{Bjerrum-Bohr:2010pnr}
N.~E.~J. Bjerrum-Bohr, P.~H. Damgaard, T.~Sondergaard, and P.~Vanhove, ``{The
  Momentum Kernel of Gauge and Gravity Theories},''
  \href{http://dx.doi.org/10.1007/JHEP01(2011)001}{{\em JHEP} {\bfseries 01}
  (2011) 001}, \href{http://arxiv.org/abs/1010.3933}{{\ttfamily arXiv:1010.3933
  [hep-th]}}.

\bibitem{Bautista:2019tdr}
Y.~F. Bautista and A.~Guevara, ``{From Scattering Amplitudes to Classical
  Physics: Universality, Double Copy and Soft Theorems},''
  \href{http://arxiv.org/abs/1903.12419}{{\ttfamily arXiv:1903.12419
  [hep-th]}}.

\bibitem{Bern:2019crd}
Z.~Bern, C.~Cheung, R.~Roiban, C.-H. Shen, M.~P. Solon, and M.~Zeng, ``{Black
  Hole Binary Dynamics from the Double Copy and Effective Theory},''
  \href{http://dx.doi.org/10.1007/JHEP10(2019)206}{{\em JHEP} {\bfseries 10}
  (2019) 206}, \href{http://arxiv.org/abs/1908.01493}{{\ttfamily
  arXiv:1908.01493 [hep-th]}}.

\bibitem{DiVecchia:2015srk}
P.~Di~Vecchia, R.~Marotta, and M.~Mojaza, ``{Soft Theorems from String
  Theory},'' \href{http://dx.doi.org/10.1002/prop.201500068}{{\em Fortsch.
  Phys.} {\bfseries 64} (2016) 389--393},
  \href{http://arxiv.org/abs/1511.04921}{{\ttfamily arXiv:1511.04921
  [hep-th]}}.

\bibitem{Sen:2017xjn}
A.~Sen, ``{Soft Theorems in Superstring Theory},''
  \href{http://dx.doi.org/10.1007/JHEP06(2017)113}{{\em JHEP} {\bfseries 06}
  (2017) 113}, \href{http://arxiv.org/abs/1702.03934}{{\ttfamily
  arXiv:1702.03934 [hep-th]}}.

\bibitem{DiVecchia:2019kle}
P.~Di~Vecchia, R.~Marotta, and M.~Mojaza, ``{Multiloop soft theorem of the
  dilaton in the bosonic string},''
  \href{http://dx.doi.org/10.1103/PhysRevD.100.041902}{{\em Phys. Rev. D}
  {\bfseries 100} no.~4, (2019) 041902},
  \href{http://arxiv.org/abs/1907.01036}{{\ttfamily arXiv:1907.01036
  [hep-th]}}.

\end{thebibliography}\endgroup

 \end{document}